\documentclass[twocolumn]{aastex631}
\usepackage{amsmath,amssymb}
\usepackage{comment}
\usepackage {autobreak}

\shorttitle{New $Y_\mathrm{P}$ Determination with EMPGs}
\shortauthors{Matsumoto et al.}
\graphicspath{{./}{figures/}}

\begin{document}

\title{
EMPRESS. VIII.\\
A New Determination of Primordial He Abundance with Extremely Metal-Poor Galaxies:\\ 
A Suggestion of the Lepton Asymmetry and Implications for the Hubble Tension
}

\author{Akinori Matsumoto}
\affiliation{Institute for Cosmic Ray Research, The University of Tokyo, 5-1-5 Kashiwanoha, Kashiwa, Chiba 277-8582, Japan}
\affiliation{Department of Physics, Graduate School of Science, The University of Tokyo, 7-3-1 Hongo, Bunkyo, Tokyo 113-0033, Japan}

\author[0000-0002-1049-6658]{Masami Ouchi}
\affiliation{National Astronomical Observatory of Japan, National Institutes of Natural Sciences, 2-21-1 Osawa, Mitaka, Tokyo 181-8588, Japan}
\affiliation{Institute for Cosmic Ray Research, The University of Tokyo, 5-1-5 Kashiwanoha, Kashiwa, Chiba 277-8582, Japan}
\affiliation{Kavli Institute for the Physics and Mathematics of the Universe (WPI), University of Tokyo, Kashiwa, Chiba 277-8583, Japan}

\author[0000-0003-2965-5070]{Kimihiko Nakajima}
\affiliation{National Astronomical Observatory of Japan, National Institutes of Natural Sciences, 2-21-1 Osawa, Mitaka, Tokyo 181-8588, Japan}

\author{Masahiro Kawasaki}
\affiliation{Institute for Cosmic Ray Research, The University of Tokyo, 5-1-5 Kashiwanoha, Kashiwa, Chiba 277-8582, Japan}
\affiliation{Kavli Institute for the Physics and Mathematics of the Universe (WPI), University of Tokyo, Kashiwa, Chiba 277-8583, Japan}

\author{Kai Murai}
\affiliation{Institute for Cosmic Ray Research, The University of Tokyo, 5-1-5 Kashiwanoha, Kashiwa, Chiba 277-8582, Japan}

\author{Kentaro Motohara}
\affiliation{National Astronomical Observatory of Japan, National Institutes of Natural Sciences, 2-21-1 Osawa, Mitaka, Tokyo 181-8588, Japan}
\affiliation{Institute of Astronomy, Graduate School of Science, The University of Tokyo, 2-21-1 Osawa, Mitaka, Tokyo 181-0015, Japan}

\author[0000-0002-6047-430X]{Yuichi Harikane}
\affiliation{Institute for Cosmic Ray Research, The University of Tokyo, 5-1-5 Kashiwanoha, Kashiwa, Chiba 277-8582, Japan}
\affiliation{Department of Physics and Astronomy, University College London, Gower Street, London WC1E 6BT, UK}

\author[0000-0001-9011-7605]{Yoshiaki Ono}
\affiliation{Institute for Cosmic Ray Research, The University of Tokyo, 5-1-5 Kashiwanoha, Kashiwa, Chiba 277-8582, Japan}

\author{Kosuke Kushibiki}
\affiliation{Institute of Astronomy, Graduate School of Science, The University of Tokyo, 2-21-1 Osawa, Mitaka, Tokyo 181-0015, Japan}

\author{Shuhei Koyama}
\affiliation{Institute of Astronomy, Graduate School of Science, The University of Tokyo, 2-21-1 Osawa, Mitaka, Tokyo 181-0015, Japan}

\author[0000-0002-1005-4120]{Shohei Aoyama}
\affiliation{Institute for Cosmic Ray Research, The University of Tokyo, 5-1-5 Kashiwanoha, Kashiwa, Chiba 277-8582, Japan}

\author{Masahiro Konishi}
\affiliation{Institute of Astronomy, Graduate School of Science, The University of Tokyo, 2-21-1 Osawa, Mitaka, Tokyo 181-0015, Japan}

\author{Hidenori Takahashi}
\affiliation{Institute of Astronomy, Graduate School of Science, The University of Tokyo, 2-21-1 Osawa, Mitaka, Tokyo 181-0015, Japan}
\affiliation{Kiso Observatory, the University of Tokyo, 10762-30, Mitake, Kiso-machi, Kiso-gun, Nagano 397-0101, Japan}

\author[0000-0001-7730-8634]{Yuki Isobe}
\affiliation{Institute for Cosmic Ray Research, The University of Tokyo, 5-1-5 Kashiwanoha, Kashiwa, Chiba 277-8582, Japan}
\affiliation{Department of Physics, Graduate School of Science, The University of Tokyo, 7-3-1 Hongo, Bunkyo, Tokyo 113-0033, Japan}

\author{Hiroya Umeda}
\affiliation{Institute for Cosmic Ray Research, The University of Tokyo, 5-1-5 Kashiwanoha, Kashiwa, Chiba 277-8582, Japan}
\affiliation{Department of Physics, Graduate School of Science, The University of Tokyo, 7-3-1 Hongo, Bunkyo, Tokyo 113-0033, Japan}

\author[0000-0001-6958-7856]{Yuma Sugahara}
\affiliation{National Astronomical Observatory of Japan, National Institutes of Natural Sciences, 2-21-1 Osawa, Mitaka, Tokyo 181-8588, Japan}
\affiliation{Waseda Research Institute for Science and Engineering, Faculty of Science and Engineering, Waseda University, 3-4-1, Okubo, Shinjuku, Tokyo 169-8555, Japan}
\affiliation{Institute for Cosmic Ray Research, The University of Tokyo, 5-1-5 Kashiwanoha, Kashiwa, Chiba 277-8582, Japan}
\affiliation{Department of Physics, Graduate School of Science, The University of Tokyo, 7-3-1 Hongo, Bunkyo, Tokyo 113-0033, Japan}

\author[0000-0003-3228-7264]{Masato Onodera}
\affiliation{Subaru Telescope, National Astronomical Observatory of Japan, National Institutes of Natural Sciences (NINS), 650 North Aohoku Place, Hilo, HI 96720, USA}
\affiliation{Department of Astronomical Science, SOKENDAI (The Graduate University for Advanced Studies), Osawa 2-21-1, Mitaka, Tokyo, 181-8588, Japan}

\author[0000-0001-7457-8487]{Kentaro Nagamine}
\affiliation{Theoretical Astrophysics, Department of Earth and Space Science, Graduate School of Science, Osaka University, 1-1 Machikaneyama, Toyonaka, Osaka 560-0043, Japan}
\affiliation{Kavli Institute for the Physics and Mathematics of the Universe (WPI), University of Tokyo, Kashiwa, Chiba 277-8583, Japan}
\affiliation{Department of Physics \& Astronomy, University of Nevada, Las Vegas, 4505 S. Maryland Pkwy, Las Vegas, NV 89154-4002, USA}

\author[0000-0002-3801-434X]{Haruka Kusakabe}
\affiliation{Observatoire de Gen\`{e}ve, Universit\'e de Gen\`{e}ve, 51 Chemin de P\'egase, 1290 Versoix, Switzerland}

\author[0000-0002-5661-033X]{Yutaka Hirai}
\affiliation{Department of Physics and Astronomy, University of Notre Dame, 225 Nieuwland Science Hall, Notre Dame, IN 46556, USA}
\affiliation{Astronomical Institute, Tohoku University, 6-3 Aoba, Aramaki, Aoba-ku, Sendai, Miyagi 980-8578, Japan}

\author[0000-0003-1169-1954]{Takashi J. Moriya}
\affiliation{National Astronomical Observatory of Japan, National Institutes of Natural Sciences, 2-21-1 Osawa, Mitaka, Tokyo 181-8588, Japan}
\affiliation{School of Physics and Astronomy, Faculty of Science, Monash University, Clayton, Victoria 3800, Australia}

\author{Takatoshi Shibuya}
\affiliation{Kitami Institute of Technology, 165 Koen-cho, Kitami, Hokkaido 090-8507, Japan}

\author{Yutaka Komiyama}
\affiliation{National Astronomical Observatory of Japan, National Institutes of Natural Sciences, 2-21-1 Osawa, Mitaka, Tokyo 181-8588, Japan}

\author{Keita Fukushima}
\affiliation{Theoretical Astrophysics, Department of Earth and Space Science, Graduate School of Science, Osaka University, 1-1 Machikaneyama, Toyonaka, Osaka 560-0043, Japan}


\author[0000-0001-7201-5066]{Seiji Fujimoto}
\affiliation{Cosmic DAWN Center}
\affiliation{Niels Bohr Institute, University of Copenhagen, Lyngbyvej2, DK-2100, Copenhagen, Denmark}
\affiliation{Waseda Research Institute for Science and Engineering, Faculty of Science and Engineering, Waseda University, 3-4-1, Okubo, Shinjuku, Tokyo 169-8555, Japan}
\affiliation{National Astronomical Observatory of Japan, National Institutes of Natural Sciences, 2-21-1 Osawa, Mitaka, Tokyo 181-8588, Japan}
\affiliation{Institute for Cosmic Ray Research, The University of Tokyo, 5-1-5 Kashiwanoha, Kashiwa, Chiba 277-8582, Japan}

\author{Takashi Hattori}
\affiliation{Subaru Telescope, National Astronomical Observatory of Japan, National Institutes of Natural Sciences (NINS), 650 North Aohoku Place, Hilo, HI 96720, USA}

\author[0000-0002-8758-8139]{Kohei Hayashi}
\affiliation{Astronomical Institute, Tohoku University, 6-3 Aoba, Aramaki, Aoba-ku, Sendai, Miyagi 980-8578, Japan}
\affiliation{National Institute of Technology, Ichinoseki College, Takanashi, Hagisho, Ichinoseki, Iwate, 021-8511, Japan}

\author[0000-0002-7779-8677]{Akio K. Inoue}
\affiliation{Waseda Research Institute for Science and Engineering, Faculty of Science and Engineering, Waseda University, 3-4-1, Okubo, Shinjuku, Tokyo 169-8555, Japan}
\affiliation{Department of Physics, School of Advanced Science and Engineering, Faculty of Science and Engineering, Waseda University, 3-4-1 Okubo, Shinjuku, Tokyo 169-8555, Japan}

\author[0000-0003-2449-6314]{Shotaro Kikuchihara}
\affiliation{Institute for Cosmic Ray Research, The University of Tokyo, 5-1-5 Kashiwanoha, Kashiwa, Chiba 277-8582, Japan}
\affiliation{Department of Astronomy, Graduate School of Science, The University of Tokyo, 7-3-1 Hongo, Bunkyo, Tokyo 113-0033, Japan}

\author[0000-0001-5780-1886]{Takashi Kojima}
\affiliation{Institute for Cosmic Ray Research, The University of Tokyo, 5-1-5 Kashiwanoha, Kashiwa, Chiba 277-8582, Japan}
\affiliation{Department of Physics, Graduate School of Science, The University of Tokyo, 7-3-1 Hongo, Bunkyo, Tokyo 113-0033, Japan}

\author[0000-0002-0479-3699]{Yusei Koyama}
\affiliation{Subaru Telescope, National Astronomical Observatory of Japan, National Institutes of Natural Sciences (NINS), 650 North Aohoku Place, Hilo, HI 96720, USA}
\affiliation{Department of Astronomical Science, SOKENDAI (The Graduate University for Advanced Studies), Osawa 2-21-1, Mitaka, Tokyo, 181-8588, Japan}

\author[0000-0003-1700-5740]{Chien-Hsiu Lee}
\affiliation{Subaru Telescope, National Astronomical Observatory of Japan, National Institutes of Natural Sciences (NINS), 650 North Aohoku
Place, Hilo, HI 96720, USA}

\author[0000-0003-4985-0201]{Ken Mawatari}
\affiliation{National Astronomical Observatory of Japan, National Institutes of Natural Sciences, 2-21-1 Osawa, Mitaka, Tokyo 181-8588, Japan}
\affiliation{Institute for Cosmic Ray Research, The University of Tokyo, 5-1-5 Kashiwanoha, Kashiwa, Chiba 277-8582, Japan}
\affiliation{Department of Environmental Science and Technology, Fuculty of Design Technology, Osaka Sangyo University, 3-1-1, Nakagaito, Daito, Osaka, 574-8530, Japan}

\author{Takashi Miyata}
\affiliation{Institute of Astronomy, Graduate School of Science, The University of Tokyo, 2-21-1 Osawa, Mitaka, Tokyo 181-0015, Japan}

\author[0000-0002-7402-5441]{Tohru Nagao}
\affiliation{Research Center for Space and Cosmic Evolution, Ehime University, Matsuyama, Ehime 790-8577, Japan}

\author[0000-0002-5443-0300]{Shinobu Ozaki}
\affiliation{National Astronomical Observatory of Japan, National Institutes of Natural Sciences, 2-21-1 Osawa, Mitaka, Tokyo 181-8588, Japan}

\author{Michael Rauch}
\affiliation{Carnegie Observatories, 813 Santa Barbara Street, Pasadena, CA 91101, USA}

\author{Tomoki Saito}
\affiliation{Nishi-Harima Astronomical Observatory, Centre for Astronomy, University of Hyogo, 407-2 Nishigaichi, Sayo, Sayo-gun, Hyogo 679-5313}

\author[0000-0002-7043-6112]{Akihiro Suzuki}
\affiliation{National Astronomical Observatory of Japan, National Institutes of Natural Sciences, 2-21-1 Osawa, Mitaka, Tokyo 181-8588, Japan}

\author{Tsutomu T. Takeuchi}
\affiliation{Department of Particle and Astrophysical Science, Nagoya University, Furo-cho, Chikusa-ku, Nagoya, 464-8602, Aichi, Japan}

\author{Masayuki Umemura}
\affiliation{Center for Computational Sciences, University of Tsukuba, Tsukuba, Ibaraki 305-8577, Japan}

\author[0000-0002-5768-8235]{Yi Xu}
\affiliation{Institute for Cosmic Ray Research, The University of Tokyo, 5-1-5 Kashiwanoha, Kashiwa, Chiba 277-8582, Japan}
\affiliation{Department of Astronomy, Graduate School of Science, The University of Tokyo, 7-3-1 Hongo, Bunkyo, Tokyo 113-0033, Japan}

\author[0000-0001-6229-4858]{Kiyoto Yabe}
\affiliation{Kavli Institute for the Physics and Mathematics of the Universe (WPI), University of Tokyo, Kashiwa, Chiba 277-8583, Japan}

\author{Yechi Zhang}
\affiliation{Institute for Cosmic Ray Research, The University of Tokyo, 5-1-5 Kashiwanoha, Kashiwa, Chiba 277-8582, Japan}
\affiliation{Department of Physics, Graduate School of Science, The University of Tokyo, 7-3-1 Hongo, Bunkyo, Tokyo 113-0033, Japan}

\author{Yuzuru Yoshii}
\affiliation{Institute of Astronomy, Graduate School of Science, The University of Tokyo, 2-21-1 Osawa, Mitaka, Tokyo 181-0015, Japan}
\affiliation{Steward Observatory, University of Arizona, 933 North Cherry Avenue, Rm. N204 Tucson, AZ 85721-0065, USA}


\begin{abstract}
The primordial He abundance $Y_\mathrm{P}$ is a powerful probe of cosmology. Currently, $Y_\mathrm{P}$ is best determined by observations of metal-poor galaxies, while there are only a few known local extremely metal-poor ($<0.1 Z_\odot$) galaxies (EMPGs) having reliable He/H measurements with He{\sc i}$\lambda$10830 near-infrared (NIR) emission. Here we present deep Subaru NIR spectroscopy for 10 EMPGs. Combining the existing optical data, He/H values of 5 out of the 10 EMPGs are reliably derived by the Markov chain Monte Carlo algorithm. Adding the existing 3 EMPGs and 51 moderately metal-poor ($0.1-0.4 Z_\odot$) galaxies with reliable He/H estimates, we obtain $Y_\mathrm{P}=0.2370^{+0.0034}_{-0.0033}$ by linear regression in the $\mathrm{(He/H)}-\mathrm{(O/H)}$ plane, where we increase the number of EMPGs from 3 to 8 anchoring He/H of the most metal-poor gas in galaxies. Although our $Y_\mathrm{P}$ measurement and previous measurements are consistent, our result is slightly ($\sim 1\sigma$) smaller due to our EMPGs. Including the existing primordial deuterium $D_\mathrm{P}$ measurement, we constrain the effective number of neutrino species $N_\mathrm{eff}$ and the baryon-to-photon ratio $\eta$ showing $\gtrsim 1-2\sigma$ tensions with the Standard Model and \citet{planck+18}. Motivated by the tensions, we allow the degeneracy parameter of electron-neutrino $\xi_e$ to vary as well as $N_\mathrm{eff}$ and $\eta$.  We obtain $\xi_e = 0.05^{+0.03}_{-0.02}$, $N_\mathrm{eff}=3.11^{+0.34}_{-0.31}$, and $\eta\times10^{10}=6.08^{+0.06}_{-0.06}$ from the $Y_\mathrm{P}$ and $D_\mathrm{P}$ measurements with a prior of $\eta$ taken from \citet{planck+18}. Our constraints suggest a lepton asymmetry and allow for a high value of $N_\mathrm{eff}$ within the $1\sigma$ level, which could mitigate the Hubble tension.
\end{abstract}


\section{Introduction} \label{sec:intro}
The flat $\Lambda$CDM model
shows good consistency with the independent observational measurements of the cosmic microwave background (CMB) \citep{planck+18}, the large scale structures, and the expansion history of the universe. However, as the precision of observations increases, a significant discrepancy between the determinations of the Hubble parameter ($H_0$) is revealed. For example, \citet{riess2019} demonstrate that the value of the direct $H_0$ measurement with 70 Cepheids is higher with $4.4\sigma$ tension than the value inferred from the Planck measurements with the $\Lambda$CDM model. This tensoin is called ``Hubble tension", and the recent studies claim $\gtrsim 5\sigma$ 
differences \citep{wong2020,riess2021}. The Hubble tension may be interpreted as evidence for new cosmological features beyond the $\Lambda$CDM model \citep[e.g.,][]{renk2017,khosaravi2019,dainotti2021,dainotti2022}.
One possible way of resolving the Hubble tension problem is to allow the effective number of neutrino species $N_\mathrm{eff}$, which can be regarded as a parameter for the total energy density of relativistic particles, 
to change. A value of $N_\mathrm{eff}$ larger than the one predicted by the Standard Model, 3.046, increases the $H_0$ value inferred from the Planck CMB observations in the $\Lambda$CDM model, reducing the scale of the sound horizon.
\citet{bernal2016} claim that $N_\mathrm{eff}\sim 3.4$ can ameliorate the Hubble tension. Similarly, \citet{vagnozzi2020} shows that the Hubble tension would be reduced to $1.5\sigma$ by models with $N_\mathrm{eff}\simeq 3.45$ and these models are only weakly disfavored compared with the standard $\Lambda$CDM model. 

Besides the Hubble tension, the determination of the value of $N_\mathrm{eff}$ is important in particle physics and cosmology. Many particle physics beyond the Standard Model and inflation models predict the existence of extra-radiation like dark radiation and gravitational waves \citep[e.g.,][]{dunsky2020}. Because $N_\mathrm{eff}$ changes from 3.046 with the presence of such extra-radiations, the measurement of $N_\mathrm{eff}$ places constraints on the extended models.
Around the epoch of the Big Bang Nucleosynthsis \citep[BBN, see e.g.,][for reviews]{BBN_2007,BBN_summary}, the radiation energy density (i.e., $N_\mathrm{eff}$) primarily determines the expansion rate of the universe via the Friedmann equation. The competition between the expansion rate and the weak interaction rate determines the ``freeze-out" value of neutron to proton abundance ratio. The neutron abundance is then reduced by free decay until BBN occurs. Since virtually all remaining neutrons are processed into $^4$He (hereafter He), the primordial He abundance in mass fraction, $Y_\mathrm{P}$,
%
%
%
offers a strong constraint on $N_\mathrm{eff}$.

Although the CMB measurements of \citet{planck+18} provide $Y_\mathrm{P}=0.246\pm0.035$ $(95\%)$, this accuracy is not good enough to determine $N_\mathrm{eff}$ with an uncertainty less than $\Delta N_\mathrm{eff}\simeq 0.6$.
The $Y_\mathrm{P}$ value can be more strongly constrained by observations for He abundances of metal-poor galaxies (i.e., galaxies whose elemental compositions are close to the primordial one), reaching sub-percent level accuracy \citep{izotov2014,aver2015,peimbert2016,valerdi2019,fernandez2019,hsyu2020,kurichin2021}. 
To derive He abundances of metal-poor galaxies for $Y_\mathrm{P}$ determination, one needs to constrain physical parameters of ionized nebulae of the galaxies (e.g., metallicity, electron density, and ionization parameter) with observed emission lines by comparisons of photoionization models. Because it is known that optical emission lines do not allow us to resolve the degeneracy between the electron density and temperature of the nebula, the near-infrared (NIR) He{\sc i}$\lambda$10830 line, which is sensitive to the electron density, is key to removing systematic uncertainties raised by the degeneracy \citep{izotov2014,aver2015}.
A recent study of metal-poor galaxies \citep{hsyu2020} uses a moderately large sample of 54 metal-poor galaxies across a metallicity range of $\mathrm{(O/H)}\times 10^{10}=1.73 - 16.64$, some of which include He{\sc i}$\lambda$10830 measurements, and report $Y_\mathrm{P}=0.2436^{+0.0036}_{-0.0040}$.
\citet{hsyu2020} obtain $N_\mathrm{eff}=2.85^{+0.28}_{-0.25}$, combining the value of $Y_\mathrm{P}$ with the measurement of primordial D to H abundance ratio $D_\mathrm{P}$ presented in \citet{cooke2018}. While the best-estimated value, $N_\mathrm{eff}=2.85^{+0.28}_{-0.25}$ is lower than $N_\mathrm{eff}=3.046$, this estimation allows $N_\mathrm{eff}=3.4$ within $2\sigma$. The uncertainty is not small enough to test whether $N_\mathrm{eff}$ can be as large as the one alleviating the Hubble tension.
Although the number of the available metal-poor galaxies with reliable He abundance measurements are moderately large, $\sim 50$, in the previous study \citep{hsyu2020},
the previous study could use only 3 galaxies of the low-metallicity end, extremely metal-poor galaxies (EMPGs) with metallicities less than 10 \% solar oxygen abundance, where the definition of the solar metallicity is given by $12+\log\mathrm{\left(O/H\right)}=8.69$ \citep{asplund2009}.
Because EMPGs possess gas of nebulae whose He abundance is much similar to the primordial He abundance compared to more metal enriched galaxies with a $>10$\% solar oxygen abundance, adding EMPGs to the sample would strongly impact on the determination of primordial He abundance.

\citet{kojima2020} have initiated a new EMPG survey named ``Extremely Metal-Poor Representatives Explored by the Subaru Survey (EMPRESS)". Having successful results of EMPRESS 
(\citealt{kojima2020,empress2,empress3,empress4,empress5,empress6,empress7}),
we have launched an extended project, EMPRESS 3D (PI: M. Ouchi) that perform optical integral-field spectroscopy (IFS) and $Y$ - band spectroscopy for $\sim 30$ EMPGs. The EMPRESS 3D project will provide deep optical spectra via the IFS data cube and weak emission lines including He{\sc i}$\lambda$10830 at the EMPG luminousity peaks.
The goals of this study are to determine $Y_\mathrm{P}$ with a high accuracy on the basis of a galaxy sample including significantly large number of EMPGs and to evaluate $N_\mathrm{eff}$ that may solve the Hubble tension.
The structure of this paper is as follows.
In Section \ref{sec:sample_data}, we present our galaxy sample. In Section \ref{sec:NIR_spec_data_reduction}, our observations and data reduction are described. We explain the data analysis and He abundance measurements of the observed galaxies in Section \ref{sec:analyses}. In Section \ref{sec:yp}, we present our determination of $Y_\mathrm{P}$ by the liner regression method. In Section \ref{sec:discussion}, we discuss the possibility of new physics beyond the standard model of cosmology. Section \ref{sec:sum} summarizes our results. 
%

\section{Sample and Data}\label{sec:sample_data}
We use a total of 64 galaxies, including 13 EMPGs, which have optical line measurements necessary for the He abundance determinations. Our sample of 64 galaxies consists of the 10 galaxies whose NIR spectra are taken by our Subaru observations (Section \ref{sec:our_taget_object}) and 54 galaxies from a sample of a previous study (Section \ref{sec:objects_literature}). In this paper, the 10 and 54 galaxies are referred to as the `Subaru galaxies' and the `literature galaxies', respectively.

\subsection{Subaru Galaxies}\label{sec:our_taget_object}

We select the Subaru galaxies from the known EMPGs whose He abundance
has not been determined with NIR data. We choose the 11 galaxies, classified as EMPGs, with $(\mathrm{O/H})=1-10\%$ solar abundance reported in previous studies (\citealt{kojima2020,empress6,izotov2012,thuan2005,papaderos2008,izotov2019,empress5}), which are bright and visible in our observing runs in January, February, April, May, and July.
%
%
The Subaru galaxies are summarized in Table \ref{tab:targets}.



\subsection{Literature Galaxies}
\label{sec:objects_literature}
For the literature galaxies, we use the 54 galaxies, 8 of which have the NIR spectroscopic data. These 54 galaxies are taken from the Sample 1 of \citet{hsyu2020} that is a sample with the reliable {\sc Hi} and He{\sc i} emission modeling. In the literature galaxies, 3 out of 54 galaxies are EMPGs. The addition of the Subaru galaxies more than quadruples the number of EMPGs for the $Y_\mathrm{P}$ determination in the previous study.
%
%
%
%


\begin{deluxetable*}{ccccc}
\tablecaption{Subaru Galaxies \label{tab:targets}}
\tablewidth{0pt}
\tablehead{
\colhead{ID} & \colhead{RA} & \colhead{Dec} & \colhead{z} &
\colhead{Reference for Optical Spectra}
}
\decimalcolnumbers
\startdata
J1631+4426 & 247.8093333 & 44.4345639 & 0.0230 &\citet{kojima2020} \\
J1418+3752 & 214.7130000 & 21.0443722 & 0.0090 &\citet{empress6}\\
J1016+3754 & 154.1022083 & 37.9127694 & 0.0039 & SDSS\\
I Zw 18 NW & 143.5084380 & 55.2411310 & 0.0024 & \citet{thuan2005}\\
J1201+0211 & 180.3430000 & 2.1856900 & 0.0030 & SDSS\\
J1119+5130 & 169.8930000 & 51.5034000 & 0.0040 & SDSS\\
J1234+3901 & 188.5654170 & 39.0212250 & 0.13297 & SDSS\\
J0133+1342 & 23.4690000 & 13.7026000 & 0.00879 & SDSS\\
J0825+3532 & 126.4810000 & 35.5422000 & 0.0020 & SDSS\\
J0125+0759 & 21.3924567 & 7.9901917 & 0.0100 & \citet{empress5}\\
J0935-0115 & 143.9133478 & -1.2615025 & 0.0162 & \citet{empress5}\\
\enddata
\tablecomments{(1): ID. (2): Right ascension. (3): Declination. (4): Redshift. (5): Reference for optical spectra.}
\end{deluxetable*}

\begin{deluxetable*}{ccccc}
\tablecaption{Near-Infrared Spectroscopy \label{tab:observation}}
\tablewidth{0pt}
\tablehead{
\colhead{ID} & \colhead{Instrument} & \colhead{Exposure Time (s)} & \colhead{Seeing (arcsec)} & \colhead{Observation Date}
}
\decimalcolnumbers
\startdata
J1631+4426 & MOIRCS & 3600 & 0.6 & 2020 July 23\\
J1418+3752 & IRCS & 1200 & 0.9 & 2021 March 31 \\
J1016+3754 & IRCS & 1200 & 0.9 & 2021 March 31 \\
I Zw 18 NW & SWIMS & 1200 & 0.4 & 2021 May 28\\
J1201+0211 & SWIMS & 1200 & 0.4 & 2021 May 28 \\
J1119+5130 & SWIMS & 1200 & 0.4 & 2021 May 28\\
J1234+3901 & SWIMS & 1200 & 0.4 & 2021 May 28\\
J0133+1342 & SWIMS & 1800 & 0.4 & 2022 January 12\\
J0825+3532 & SWIMS & 1800 & 0.4 & 2022 January 12\\
J0125+0759 & SWIMS & 1800 & 0.4 & 2022 January 12\\
J0935-0115 & SWIMS & 720 & 0.4 & 2022 February 8\\
\enddata
\tablecomments{(1): ID. (2): Instruments for our NIR spectroscopy. (3): Total exposure time. (4): FWHM of the seeing size. (5): Date of our NIR spectroscopy.}
\end{deluxetable*}

\section{NIR Spectroscopy and Data Reduction} \label{sec:NIR_spec_data_reduction}
We observed the Subaru galaxies with three NIR spectrographs on the Subaru telescope,
Multi-Object Infrared Camera and Spectrograph \citep[MOIRCS;][]{ichikawa2006,suzuki2008},
Infrared Camera and Spectrograph \citep[IRCS;][]{tokunaga1998,kobayashi2000}, and Simultaneous-color Wide-field Infrared Multi-object Spectrograph \citep[SWIMS;][]{motohara2014,motohara2016,konishi2018,konishi2020}. These observations are summarized in Table \ref{tab:observation}.

\subsection{MOIRCS}
\subsubsection{MOIRCS Observations}
The NIR spectroscopy for one of the Subaru galaxies, J1631+4426, was conducted using MOIRCS on the date of 2020 July 23 with the $zJ500$ grism and a $0\farcs8$ wide long slit, yielding spectra spanning $0.9 - 1.78~\mathrm{\mu m}$. The resolving power was $R \simeq 300$. Dome flats were obtained at the beginning of the night. We moved the telescope in an AB dithering pattern with the total exposure time of 1800 seconds. We took the spectrum of a standard star, HIP89634, for our flux calibration.

\subsubsection{MOIRCS Data Reduction}
Data reduction is performed with the IRAF package. The reduction and calibration processes include flat fielding, cosmic ray cleaning, wavelength calibration, background subtraction, and combining the nod positions before one-dimensional spectrum extraction. Wavelength solutions for MOIRCS spectra are obtained from the ThAr lamp. We then extract one-dimensional spectra and calibrate fluxes.
We extract one-dimensional spectra with a boxcar aperture that encompasses roughly 99\% of the emission. We extract the error spectra considering read-out noise and photon noise of sky and object emission, and the uncertainty of the flux calibrations. The last source of error is accounted for assuming a $2\%$ relative flux uncertainty based on observations of standard stars \citep{oke1990}. Figure \ref{fig:spectra} shows the one-dimensional spectra taken with MOIRCS.

\subsection{IRCS}
\subsubsection{IRCS Observations}
We carried out near-infrared spectroscopy for two of the Subaru galaxies, J1418+2102 and J1016+3754, with IRCS on 2021 March 31
\footnote{
Although we observed a galaxy J1253-0312 with IRCS, after the IRCS observations we recognized that J1253-0312 was not classified as an EMPG, which had an oxygen abundance of $25\%$ the solar abundance.}.
We used the $zJ$ grism with the 52 mas pixel scale with an observed-wavelength coverage of approximately $1.03 - 1.18~\mathrm{\mu m}$. 
The spectral resolution of $R \simeq 300$ was accomplished with a $0\farcs6$ wide long slit.
We took dome flats at the beginning of the night.
We performed ABBA dithering with an individual exposure of 300 seconds. We observed an A0V Hipprcos star, HIP68868, at an airmass similar to those of our targets for flux calibration.


\subsubsection{IRCS Data Reduction}
The IRCS spectra and the error spectra are processed in the same manner as the MOIRCS spectrum.
Note that the emission lines of the IRCS spectra have
profiles similar to rectangular shapes (see Figure \ref{fig:spectra}).
This is because we observed spatially extended objects with 
a large slit width
for the IRCS instrument.
The IRCS spectra are convolutions of the instrumental profile with a box shape of the slit.
These spectral shapes are commonly found in the IRCS spectra for the similarly extended targets and the observational configuration, though line flux measurements are not affected.
%

\subsection{SWIMS}
\subsubsection{SWIMS Observations}

We conducted NIR spectroscopy with SWIMS for eight of the Subaru galaxies, I Zw 18 NW, J1201+0211, J1119+5130, and J1234+3901 on 2021 May 28, J0133+1342, J0825+3532, and J0125+0759 on 2022 January 12, and J0935-0115 on 2022 February 8. We utilized the multi-object spectroscopy mode and long-slit spectroscopy mode for J0125+0759 and the other galaxies, respectively.
%
%
%
%
The $zJ$ and $HK_{\rm s}$ grisms were used with the blue and red channels, respectively, with the dichroic at $1.4~\mathrm{\mu m}$, resulting in an observed-wavelength coverage of approximately $0.9 - 2.5~\mathrm{\mu m}$. With a slit width of $0^{\prime\prime}.8$, 
the spectral resolutions were $R \sim 700-1200$ and $600-1000$ in the blue and red channels, respectively.
Dome flats were obtained at the beginning of the nights. We used an ABBA dither pattern with individual exposures of 300 seconds for the targets except for J0125+0759. The individual exposure time was 180 seconds for J0125+0759.
For flux calibration, an A0V Hipparcos stars, HIP59861, HIP116886, and HIP19578 were observed 
at an airmass similar to the one of our targets on the dates of 2021 May 28, 2022 January 12, and February 8 respectively.


\subsubsection{SWIMS Data Reduction}
\begin{figure*}
\gridline{\fig{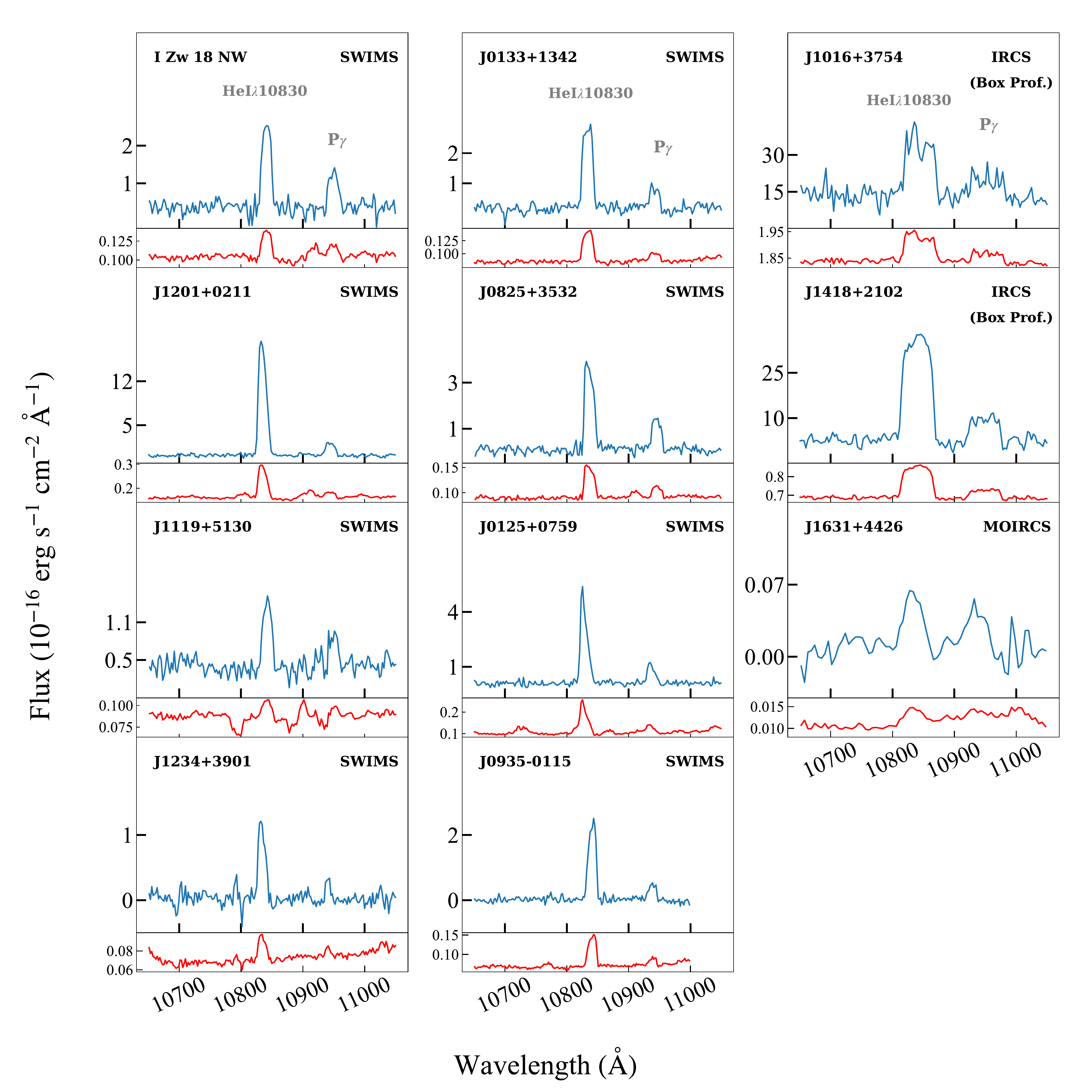}{1\textwidth}{}}
\caption{Rest-frame spectra of Subaru galaxies, taken with IRCS, SWIMS, and MOIRCS.
The blue lines and red lines present the fluxes and the $1\sigma$ errors, respectively.
%
%
Note that 
%
the emission lines of the IRCS spectra have
profiles similar to box shapes
due to 
the spatially-extended objects observed with the large slit width for IRCS,
while this does not affect the flux measurements by summing the pixels above the continuum.
%
\label{fig:spectra}}
\end{figure*}

We reduce the SWIMS spectra and the error spectra in the same manner as the MOIRCS and the IRCS data reduction.
Because the second spectrum obtained at the nod location B of each ABBA dither on 2021 May 28 includes systematic uncertainties due to a SWIMS instrument problem, we remove their second spectra in the ABBA dither data sets.
We obtain wavelength solutions for the SWIMS spectra from OH sky lines. The reduced SWIMS spectra are presented in Figure \ref{fig:spectra}.

\section{Analyses}
\label{sec:analyses}
\subsection{Flux and EW Measurements}
\label{sec:measurements}

We measure hydrogen, helium, oxygen, and sulfur emission line fluxes.
While the different instruments are used for the optical and NIR spectroscopy, we assume that the same region of each galaxy was observed. Under the assumption, the difference in aperture does not affect our analysis because we measure the ratios of the optical (NIR) line fluxes to H$\beta$ (P$\gamma$) fluxes to avoid systematics caused by the different amount of slit-loss fluxes.


\subsubsection{Optical Spectra}
\label{sec:opt_meaesurements}
We use the optical spectra and the corresponding error spectra of Magellan/MagE, Keck/DEIMOS, and SDSS for the Subaru galaxies obtained by previous observations (see Table \ref{tab:targets}) for our optical line measurements. The exception is I Zw 18 NW, for which we use the emission line flux values reported in \citet{thuan2005} with higher precision than that of the SDSS spectrum. For the Subaru galaxies except for I Zw 18 NW, we define the continuum by fitting a polynomial of degree 3 to the range that are deemed by visual inspection to be free of any emission and absorption lines. After subtraction of the  continuum from the spectra, we measure line fluxes and EWs of the optical emission lines, {\sc [Oii]}$\lambda$3727, He{\sc i}$\lambda$3889, He{\sc i}$\lambda$4026, {\sc[Oiii]}$\lambda$4363, He{\sc i}$\lambda$4471, He{\sc i}$\lambda$4686, {\sc[Oiii]}$\lambda$5007, He{\sc i}$\lambda$5015, He{\sc i}$\lambda$5876, He{\sc i}$\lambda$6678, {\sc [Sii]}$\lambda$6717, {\sc[Sii]}$\lambda$6731, He{\sc i}$\lambda$7065, {\sc [Oii]}$\lambda$7320, {\sc[Oii]}$\lambda$7330, {\sc[Siii]}$\lambda$9069, the Balmer series from H$\alpha$ to H$\delta$, and the blended H8+He{\sc i}$\lambda$3889.
However, there are residual offsets around some emission lines in the continuum-subtracted spectra, which introduces systematic uncertainties in the flux measurements. These residual offsets cannot be defined by fitting constants to the regions around the emission lines because some of the emission lines with other nearby emission lines or the stellar absorption do not have surrounding flat regions wide enough to define the offsets. As such, we simultaneously fit a Gaussian and a constant to each emission lines in the continuum-subtracted spectra, where we exclude the non-flat regions around the emission lines from the fitting ranges. In this consistent manner for all emission lines, we measure the line fluxes with the Gaussian profiles correcting for the residual offsets with the constants. The fluxes of the blended lines are calculated by fitting a double Gaussian, except for the blended H8+He{\sc i}$\lambda$3889, whose profiles are well fit by a single Gaussian. To fit the continuum and the emission lines, we use {\tt\string scipy.optimize} package, which employs a $\chi^2$ minimization approach considering the error spectra. In the error spectra from \citet{kojima2020,empress5,empress6}, for which the flux calibration was performed with only one standard star, we include the $2\%$ relative flux uncertainty of standard stars \citep{oke1990}. The error of the emission line fluxes are calculated from uncertainties of the Gaussian fit. Since one-dimensional spectra are binned by a pixel of a resolution element, the correct method of the Gaussian fit is to bin the fitting function by a pixel before fitting to the spectrum. In our optical flux measurements, however, we fit unbinned Gaussian profiles to the spectra for simplicity assuming that the choice of the fitting methods does not affect our result. We have confirmed this assumption comparing both cases of the fitting methods for one of the Subaru galaxies, J1201+0211. As we expect, the line fluxes of J1201+0211 derived by these two methods are virtually indistinguishable. For example, the H$\beta$ fluxes in units of $10^{-16}~\mathrm{erg~s^{-1} cm^{-2}}$ obtained by the Gaussian fitting with binning and without binning are $116.394 \pm 1.692$ and $116.388\pm1.729$, respectively.

The exception is J1418+3752, whose emission lines have broader profiles due to outflows. Because emission lines of J1418+3752 cannot be fit with a single Gaussian, we exclude this galaxy for our $Y_\mathrm{P}$ determination.

We also measure the line fluxes by summing the pixels above the continuum level (hereafter referred to as the integration method), but the residual offsets in the continuum-subtracted spectra turn out to introduce systematic uncertainties to the optical line flux values. We therefore adopt line fluxes measured by a Gaussian fit for the optical spectra. In Section \ref{sec:y+} and \ref{sec:yp-regression}, we discuss the impact of our choice of the methods for the line fluxes measurements on our result.


\subsubsection{NIR Spectra}
\label{sec:nir_meaesurements}
Because emission lines of MOIRCS, SWIMS, and IRCS spectra, unlike those of the optical spectra, are not well represented by a single Gaussian (see Figure \ref{fig:spectra}), the fluxes of these lines are derived by summing the pixels above the continuum determined in the same manner as the optical flux measurements, where there are no obvious residual offsets from the defined continuum of the NIR spectra. We estimate the line flux errors propagating the uncertainty of the continuum level to the error spectra.
We show the derived flux ratios and equivalent widths of the He{\sc i}$\lambda$10830 emission line of the Subaru galaxies in Table \ref{tab:fx_ratios}.

\begin{deluxetable*}{ccc}
\tablecaption{Flux Ratio and Equivalent Width of He{\sc i}$\lambda$10830 for the Subaru Galaxies \label{tab:fx_ratios}}
\tablewidth{0pt}
\tablehead{
\colhead{ID} & \colhead{$F(\mathrm{HeI\lambda10830})/F(\mathrm{P\gamma})$} & \colhead{EW (\AA)}}
\decimalcolnumbers
\startdata
J1631+4426 & $1.29\pm0.28$ & $144.1\pm 21.1$\\
J1016+3754 & $2.16\pm0.13$ & $103.0\pm3.6$\\
I Zw 18 NW & $2.18\pm0.11$ & $108.8\pm5.2$\\
J1201+0211 & $7.90\pm0.36$ & $1108.6\pm82.4$ \\
J1119+5130 & $2.02\pm0.16$ & $44.0\pm2.4$\\
J1234+3901 & $5.83\pm0.89$ & $868.1\pm406.0$\\
J0133+1342 & $4.05\pm0.26$ & $364.4\pm27.8$\\
J0825+3532 & $2.80\pm0.10$ & $790.5\pm96.8$\\
J0125+0759 & $4.21\pm0.27$ & $631.0\pm51.2$\\
J0935-0115 & $5.45\pm0.55$ & $3915.3\pm3485.3$\\
\enddata
\tablecomments{(1): ID. (2): Flux ratio of He{\sc i}$\lambda$10830 to P$\gamma$. (3): Equivalent width of the He{\sc i}$\lambda$10830 emission line.}
\end{deluxetable*}
\subsection{He Abundance}
We assume that almost all hydrogen atoms are ionized in the region considered in this paper. Therefore, the abundance ratio of helium to hydrogen $y$ is given by the sum of the abundance ratios of neutral $y^0$, singly ionized $y^+$, and doubly ionized $y^{++}$ helium to ionized hydrogen.
\begin{equation}\label{eq:y_def}
    y = y^{++} + y^+ + y^0. 
\end{equation}

We derive the $y$ values of the Subaru galaxies described in Section \ref{sec:our_taget_object}, following the procedures similar to those of \citet{hsyu2020}. The $y^{++}$ value is calculated with the He{\sc ii}$\lambda$4686 line flux. The $y^+$ value is derived based on the Markov chain Monte Carlo (MCMC) analysis with multiple He{\sc i} and H{\sc i} lines. We then check the contribution of the neutral He abundance to the $y$ value. We describe the details in Section \ref{sec:y++}, \ref{sec:y+}, and \ref{sec:yn} below. The derived $y$ values are listed in Table \ref{tab:he-o}.

\subsubsection{Doubly Ionized Helium Abundance Ratios}\label{sec:y++}
We estimate $y^{++}$ values with line flux ratios of He{\sc ii} $\lambda4686$ to H$\beta$ with the equation (17) of \citet{pagel1992}:
\begin{equation}\label{eq:y++}
    y^{++} = 0.084 t\left(\mathrm{{\sc OIII}}\right)^{0.14}\frac{F\left(\mathrm{He{\sc II}}\lambda4686\right)}{F\left(\mathrm{H}\beta\right)} 
\end{equation}
where $t\left(\mathrm{O\sc{III}}\right)$ is the electron temperature in units of $10^4~\mathrm{K}$ in the doubly ionized oxygen region. We use reported values of $t\left(\mathrm{O\sc{III}}\right)$ \citep{kojima2020,izotov2012,thuan2005,papaderos2008,izotov2019}. 
If there is no detectable He{\sc ii}$\lambda4686$ line in a galaxy, the $y^{++}$ abundance of the galaxy is assumed to be negligible. The $y^{++}$ values of the Subaru galaxies are listed in Table \ref{tab:ymcmc}.

\subsubsection{Singly Ionized Helium Abundance Ratios}\label{sec:y+}
We determine the $y^+$ values of each galaxy using the YMCMC code developed by \citet{hsyu2020}. Exploiting the MCMC algorithm, the YMCMC code conducts 
model fitting to the observed emission line ratios among He{\sc i}$\lambda$3889, He{\sc i}$\lambda$4026, He{\sc i}$\lambda$4471, He{\sc i}$\lambda$5015, He{\sc i}$\lambda$5876, He{\sc i}$\lambda$6678, He{\sc i}$\lambda$7065, the Balmer series from H$\alpha$ to H8, He{\sc i}$\lambda$10830, and P$\gamma$
to constrain 8 free parameters, $y^+,~T_\mathrm{e},~n_\mathrm{e},~ c(H\beta),~a_\mathrm{H},~a_\mathrm{He},~\tau_\mathrm{He}$, and $\xi$ with $68\%$ errors. $T_e~\mathrm{[K]}$ is the electron temperature, $n_\mathrm{e}~\mathrm{[cm^{-3}]}$ is the electron density, $c(\mathrm{H\beta})$ is the parameter of the correction for reddening, $a_\mathrm{H}~\mathrm{[\AA]}$ is the hydrogen stellar absorption normalized to that at H$\beta$, $a_\mathrm{He}~\mathrm{[\AA]}$ is the helium stellar absorption normalized to the value at He{\sc i}$\lambda4471$, $\tau_\mathrm{He}$ is the helium optical depth normalized to the value at He{\sc i}$\lambda3889$, and $\xi$ is the ratio of number density of the neutral to singly ionized hydrogen. Here, We use 500 walkers, 1000 steps, and a burn-in of 800 steps without ‘thinning’, following the procedure of \citet{hsyu2020}. For I Zw 18 NW, we do not employ the He{\sc i}$\lambda$5876 in the analysis though its line flux ratio is reported in \citet{thuan2005} because it is blended with Galactic interstellar sodium absorption \citep{izotov1999}.

In the YMCMC analysis, we assume that the observed flux ratios arise from corresponding `true' flux ratios and thus the best recovered parameters are scattered from the `true' parameter values that reproduce the `true' flux ratios. For example, if the `true' values of $c(H\beta),~a_\mathrm{H},~ a_\mathrm{He}$, and $\tau_\mathrm{He}$, which cannot originally be negative, are positive but close to 0, their best-fit values that reproduce the observed flux ratios can be negative. In this case, the expectation values of the estimated parameters we obtain restricting $c(H\beta),~a_\mathrm{H},~ a_\mathrm{He}$, and $\tau_\mathrm{He}$ to be positive can be different from the `true' values, while those extending prior boundaries can be equal to the `true' values.

\begin{figure*}
\gridline{\fig{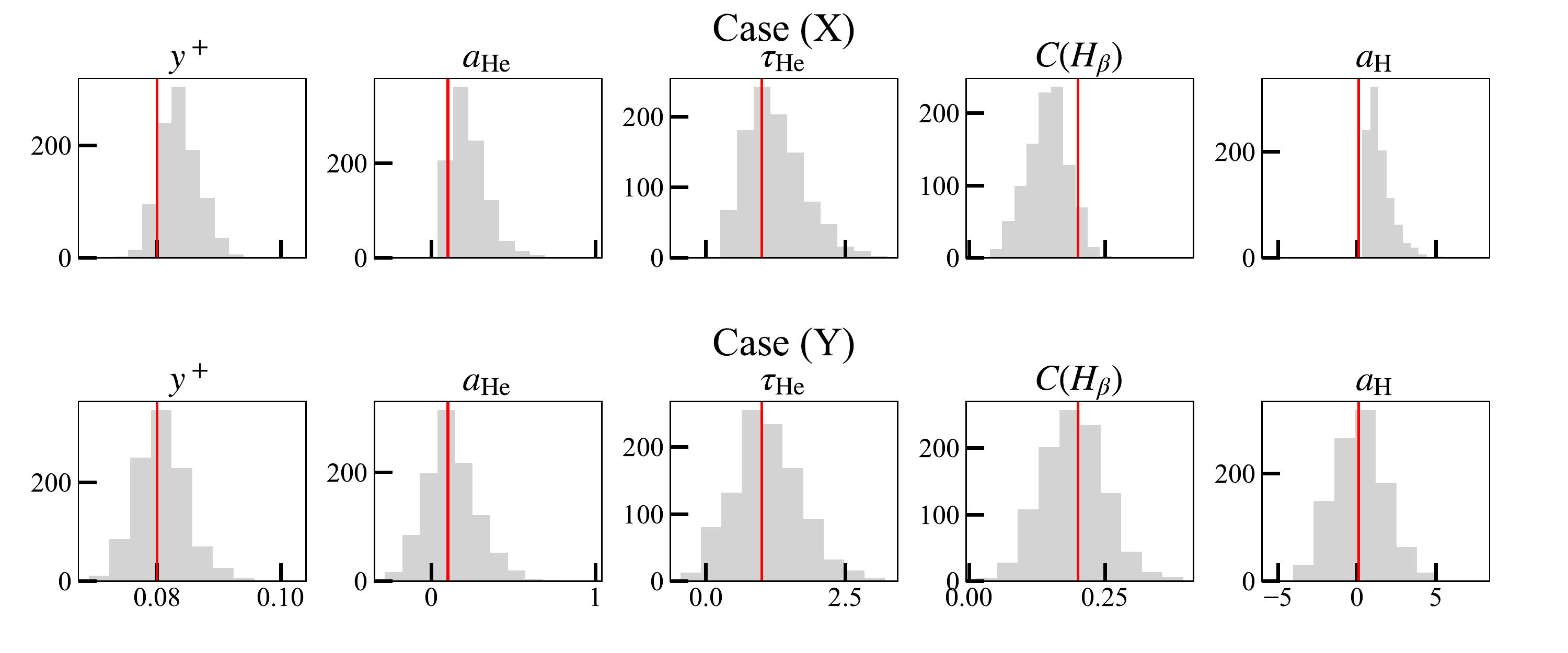}{1\textwidth}{}}
\caption{Histograms showing the distributions of the best recovered values of $y^+,~a_\mathrm{He},~\tau_\mathrm{He},~c(H\beta)$, and $a_\mathrm{H}$ for the mock `observed' flux sets in cases (X) (top panels) and (Y) (bottom panels). The solid red lines represent the `true' values of the parameters.
\label{fig:check_prior}}
\end{figure*}

To confirm the effects of the prior ranges on the expectation values of the recovered parameters, we compare simulations of the YMCMC analysis obtained in two different ways:
\begin{itemize}
\item[(X)] with the flat priors of
\begin{gather*}\label{eq:prior_hsyu}
    0.06 \leq y^+ \leq 0.10\\
    0 \leq \log_{10}(n_e) \leq 3\\
    0 \leq c(H\beta) \leq 0.5\\
    0 \leq a_\mathrm{H} \leq 10\\
    0 \leq a_\mathrm{He} \leq 4\\
    0 \leq \tau_\mathrm{He} \leq 5\\
    -6 \leq \log_{10}(\xi) \leq -0.0969
\end{gather*}
used in \citet{hsyu2020};
\item[(Y)] with the flat priors of
\begin{gather*}\label{eq:prior_ours}
    0.01 \leq y^+ \leq 0.15\\
    0 \leq \log_{10}(n_e) \leq 5\\
    -1 \leq c(H\beta) \leq 2\\
    -10 \leq a_\mathrm{H} \leq 10\\
    -5 \leq a_\mathrm{He} \leq 5\\
    -7 \leq \tau_\mathrm{He} \leq 8\\
    -10 \leq \log_{10}(\xi) \leq 0.5
\end{gather*}
,whose ranges are wide enough for this simulation to reveal the overall shapes of the probability distribution functions that are not distorted by the prior boundaries.
\end{itemize}
While the values of $c(H\beta),~a_\mathrm{H},~a_\mathrm{He}$, and $\tau_\mathrm{He}$ are restricted to be positive in case (X), they are allowed to be negative in case (Y). We generate mock `true' flux ratios by making use of the YMCMC code, in which flux ratios are predicted given the 8 physical parameters. Here, we set the `true' parameter values of $y^+=0.080,~\log_{10}n_e=2.5,~c(H\beta)=0.2,~a_\mathrm{H}=0.1,~a_\mathrm{He}=0.1,~\tau_\mathrm{He}=1,~\log_{10}(\xi)=-5$, and $T_e=18000$ to generate the mock `true' flux ratios. Note that the `true' values of $a_\mathrm{H},~a_\mathrm{He}$, and $c(H_\beta)$ are close to the lower bounds of the flat priors in case (X), $a_\mathrm{H}>0, a_\mathrm{He}>0$, and $c(H_\beta)>0$. The EW values of these fluxes, needed to run the YMCMC code, are also set. The mock flux ratios and EWs are listed in Table \ref{tab:mock_data}. Then, 1000 sets of mock `observed' fluxes are drawn from Gaussian distributions with central values of the mock `true' fluxes and scatters of $5\%$ of the central values. For each set of the mock `observed' fluxes, we derive the best-fit parameters using the YMCMC code in both cases (X) and (Y). In Figure \ref{fig:check_prior}, we show histograms for the distributions of the best-fit values of $y^+,~c(H\beta),~a_\mathrm{H},~a_\mathrm{He}$, and $\tau_\mathrm{He}$ for each set of `observed' flux ratios in cases (X) and (Y). While the best recovered values in case (X) are systematically affected, those in case (Y) are distributed around the `true' value. Because our linear regression in Section \ref{sec:yp-regression} assumes that the expectation values of the derived He/H values of galaxies are equal to their true He/H values, we extend the flat  prior ranges from those of \citet{hsyu2020} in our YMCMC analysis to prevent the parameters from being pushed up against the prior boundaries as best as possible.

\begin{figure*}
\gridline{\fig{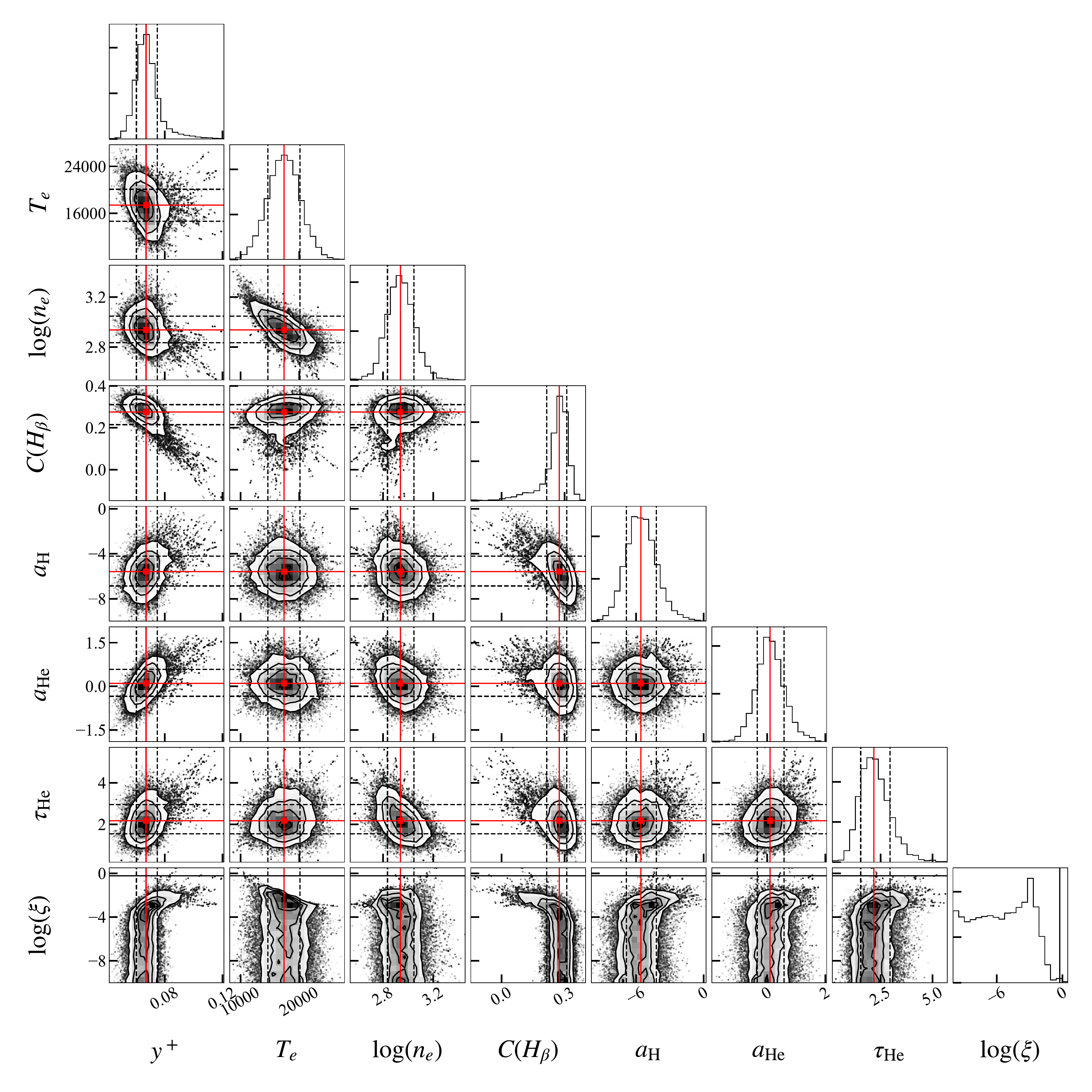}{1\textwidth}{}}
\caption{Probability distribution functions (PDFs) of model parameters of J1201+0211 recovered with YMCMC \citep{hsyu2020}. One-dimensional PDFs are shown in the diagonal panels, and the two-dimensional PDFs are shown in the off-diagonal panels with contours showing $1\sigma$, $2\sigma$, and $3\sigma$ levels. The red solid lines represent the best-recovered parameter value, and the black dashed lines represents the $68 \%$ confidence range for each parameter. In the panels showing the results for $\log \left(\xi\right)$, we show a $2\sigma$ upper limit on $\log \left(\xi\right)$ with the black solid lines.
\label{fig:eg_ymcmc}}
\end{figure*}

Using the YMCMC code, we obtain the best-fit parameters for the Subaru galaxies and present these values in Table \ref{tab:ymcmc}. As an example, we show contours and histograms for the recovered model parameters of J1201+0211 in Figure \ref{fig:eg_ymcmc}.

\begin{deluxetable*}{ccc}
\tablecaption{Mock `true' flux ratios and EWs \label{tab:mock_data}}
\tablewidth{0pt}
\tablehead{
\colhead{Ion} & \colhead{Flux Ratio} & \colhead{EW (\AA)}}
\decimalcolnumbers
\startdata
H8+He{\sc i}$\lambda$3889 & $0.1816\pm0.0091^a$ & $22.0\pm 2.2$\\
He{\sc i}$\lambda$4026 & $0.01553\pm0.00078^a$ & $2.0\pm0.2$\\
H$\delta\lambda$4101 & $0.237\pm0.012^a$ & $30.0\pm3.0$\\
H$\gamma\lambda$4340 & $0.441\pm0.022^a$ & $70.0\pm7.0$ \\
He{\sc i}$\lambda$4471 & $0.0363\pm0.0018^a$ & $6.0\pm0.6$\\
H$\beta\lambda$4861 & $1.00\pm0.05^a$ & $200.0\pm20.0$\\
He{\sc i}$\lambda$5876 & $0.1196\pm0.0060^a$ & $34.0\pm3.4$\\
H$\alpha\lambda$6563 & $4.21\pm0.27^a$ & $631.0\pm51.2$\\
He{\sc i}$\lambda$6678 & $3.17\pm0.16^a$ & $1150.0\pm115.0$\\
He{\sc i}$\lambda$7065 & $0.0399\pm0.0020^a$ & $12.0\pm1.20$\\
He{\sc i}$\lambda$10830 & $5.44\pm0.27^b$ & $380.0\pm38.0$\\
P$\gamma\lambda$10940 & $1.00\pm0.05^b$ & $160\pm16$\\
\enddata
\tablecomments{(1): Ion. (2): Flux ratio. (3): Equivalent width. \\$^a$ Normalized to the H$\beta$ flux. \\$^b$ Normalized to the P$\gamma$ flux.}
\end{deluxetable*}

To identify the galaxies whose parameters are reliably determined, for each galaxy we calculate the $\chi_\mathrm{YMCMC}^2$ value given by
\begin{equation}\label{eq:chi}
    \chi_\mathrm{YMCMC}^2 = \sum_{\lambda}\frac{\left(\frac{F(\lambda)}{F\left(\mathrm{H}\beta\right)}_\mathrm{obs}-\frac{F(\lambda)}{F\left(\mathrm{H}\beta\right)}_\mathrm{mod}\right)^2}{\sigma(\lambda)^2}
\end{equation}
where $F(\lambda)/F\left(\mathrm{H}\beta\right)$ and $\sigma(\lambda)$ are the line flux ratio and its uncertainty at wavelength $\lambda$. The subscripts obs and mod represent the observational flux ratios and the model flux ratios with the best-fit parameters the YMCMC code derives, respectively. The $\chi_\mathrm{YMCMC}^2$ value is assumed to follow the $\chi^2$-distribution with $n_\lambda - n_\mathrm{param}$ degree of freedom, where $n_\mathrm{\lambda}$ and $n_\mathrm{param}$ are the number of the measured emission lines and that of properties the YMCMC code recovers (i.e., 8), respectively.
We require that the model flux ratios with the 8 best-fit parameters are consistent with the observed flux ratios within the $95\%$ confidence level.  Among the Subaru galaxies, the 5 galaxies, J1016+3754, I Zw 18 NW, J1201+0211, J1119+5130, and J1234+3901, qualify via the $\chi^2$-criterion. In our $Y_\mathrm{P}$ determination, we use these 5 galaxies excluding the rest, J1631+4426, J0133+1342, J0825+3532, J0125+0759, and J0935-0115.

\begin{figure*}
\gridline{\fig{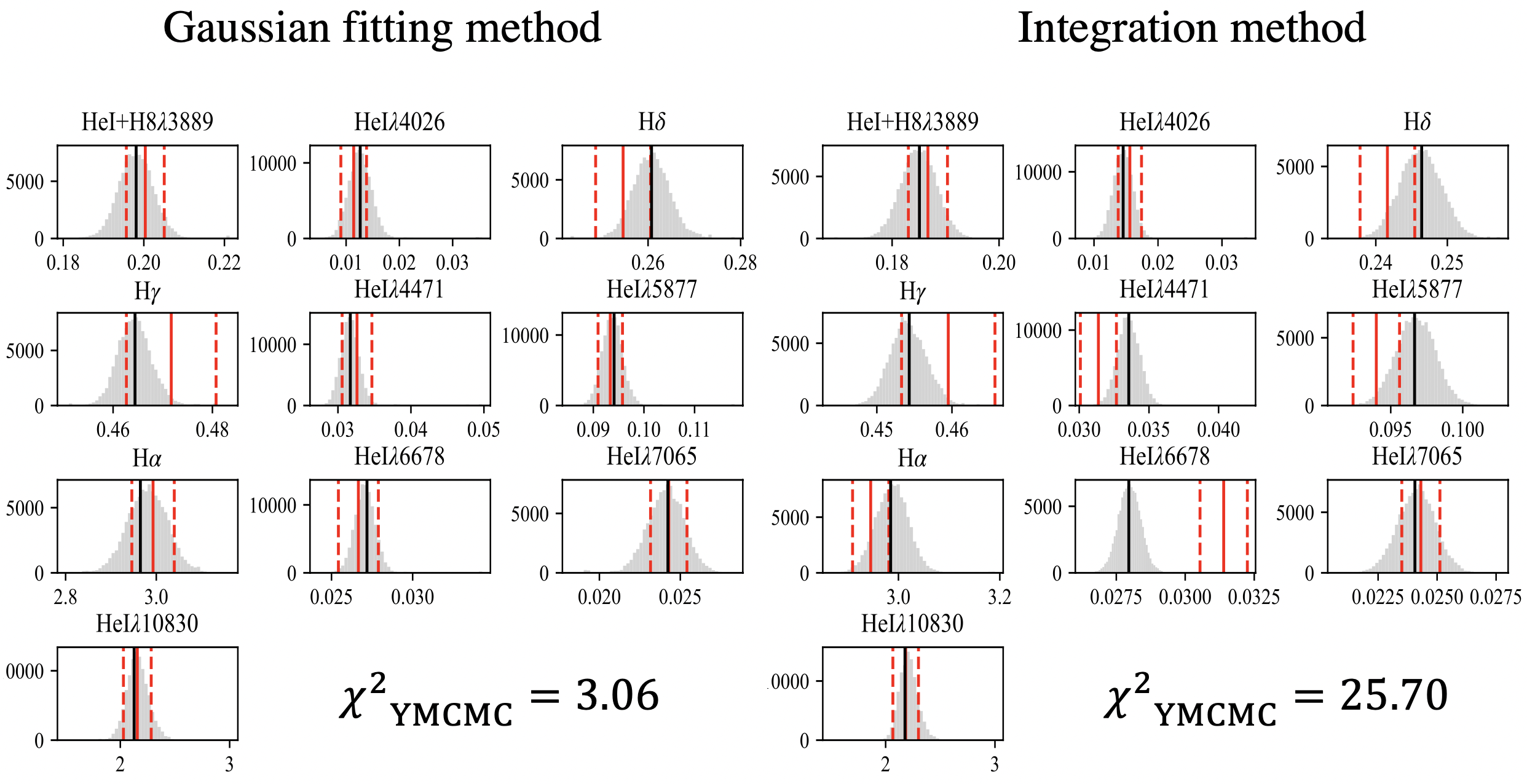}{1\textwidth}{}}
\caption{Comparison of the distributions of emission line flux ratios of J1016+3754 derived at each step of the MCMC analysis with optical fluxes of the Gaussian fitting method (left) and the integration method (right). The black solid lines show the flux ratios with the best-recovered parameters. The red solid and dashed lines show the measured line flux ratios and their $\pm 1\sigma$ values. 
\label{fig:J1016-comparison}}
\end{figure*}

To check whether our choice of flux measuring methods (Sections \ref{sec:opt_meaesurements}) has any impact on the YMCMC analysis, we solve for the best-fit parameters of the Subaru galaxies with the optical fluxes of the integration method using the YMCMC code. Applying the chi-squared criterion, we identify six galaxies (J1016+3754 and the all galaxies eliminated above), whose physical parameters are not reliably recovered with the line fluxes. We find that the galaxies tend to have the larger $\chi^2_\mathrm{YMCMC}$ values than those derived with the optical line fluxes by the method of the Gaussian fit, regardless of whether they meet the chi-squared criterion or not. As an example, Figure \ref{fig:J1016-comparison} shows histograms for the distributions of the reproduced line flux ratios of J1016+3754, derived at each step of the YMCMC analysis in both cases of the optical fluxes of the Gaussian fitting method  and the integration method together with the  $
\chi^2_\mathrm{YMCMC}$ values, $3.33$ and $25.70$, respectively. While He{\sc i}$\lambda$6678 line is reproduced within $1\sigma$ in the former case, it is not reproduced accurately (beyond the $3\sigma$ level) in the latter case, which may be attributed to the systematics made by the residual offset in the continuum-subtracted spectrum in the window of the integration. However, the derived $y$ values of the Subaru galaxies in both cases of the optical fluxes are consistent, which makes almost no difference to the $Y_\mathrm{P}$ determination (see Section \ref{sec:yp-regression}).

\begin{deluxetable*}{ccc}
\tablecaption{He and O abundance for the Subaru Galaxies \label{tab:he-o}}
\tablewidth{0pt}
\tablehead{\colhead{ID} & \colhead{$y$} & \colhead{$\mathrm{O/H}\times 10^5$~(Reference)}}
\decimalcolnumbers
\startdata
J1631+4426 & $0.0617^{+0.0101}_{-0.0094}$& $0.79\pm0.06$ \citep{kojima2020}\\
J1016+3754 & $0.0778^{+0.0034}_{-0.0027}$& $4.37\pm0.10$ \citep{izotov2012}\\
I Zw 18 NW & $0.0703^{+0.0032}_{-0.0035}$& $1.49\pm0.04$ \citep{thuan2005}\\
J1201+0211 & $0.0677^{+0.0078}_{-0.0063}$& $3.12\pm0.11$ \citep{papaderos2008}\\
J1119+5130 & $0.0810^{+0.0043}_{-0.0040}$& $3.20\pm0.17$ \citep{izotov2012}\\
J1234+3901 & $0.0804^{+0.0198}_{-0.0166}$& $1.09\pm0.07$ \citep{izotov2019}\\
J0133+1342 & $0.0777^{+0.0065}_{-0.0056}$& $3.64\pm0.11$ \citep{papaderos2008}\\
J0825+3532 & $0.0544^{+0.0142}_{-0.0048}$& $2.86\pm0.08$ \citep{thuan2005}\\
J0125+0759 & $0.0935^{+0.0096}_{-0.0055}$& $4.47\pm0.19$ \citep{empress5}\\
J0935-0115 & $0.0688^{+0.0032}_{-0.0035}$& $1.49\pm0.22$ \citep{empress5}\\
\enddata
\tablecomments{(1): ID. (2): Abundance ratio of helium to hydrogen (Equation \ref{eq:y_def}). (3): Abundance ratio of oxygen to hydrogen we use and its reference.}
\end{deluxetable*}

\begin{deluxetable*}{ccccccccccc}
\tablecaption{Physical Properties of the Subaru Galaxies  \label{tab:ymcmc}}
\tablewidth{0pt}
\tablehead{
\colhead{ID} & \colhead{$T_e$} & \colhead{$\log_{10}n_e$} & \colhead{$c(H\beta)$} & \colhead{$a_\mathrm{H}$} & \colhead{$a_\mathrm{He}$} & \colhead{$\tau_{He}$}  & \colhead{$\log_{10} (\xi)$} & \colhead{$y^+$} & \colhead{$y^{++}$} &\colhead{$\eta_{\mathrm{rds}}$} }
\decimalcolnumbers
\startdata
J1631+4426$^a$ & $24890^{+3490}_{-3600}$& $-1.39^{+1.88}_{-1.81}$& $-0.061^{+0.041}_{-0.043}$& $6.46^{+1.15}_{-1.08}$& $0.06^{+0.44}_{-0.39}$& $-5.38^{+2.97}_{-3.48}$& $-6.9^{+2.1}_{-2.1}$& $0.0596^{+0.0102}_{-0.0094}$ & $0.0020^{+0.0004}_{- 0.0004}$& $0.15^{+0.19}_{-0.23}$\\
J1016+3754 & $18400^{+1800}_{-1810}$& $0.79^{+0.49}_{-0.53}$& $0.120^{+0.028}_{-0.037}$& $-1.20^{+0.56}_{-0.51}$& $0.14^{+0.12}_{-0.11}$& $0.15^{+0.52}_{-0.50}$& $-6.0^{+2.7}_{-2.7}$& $0.0761^{+0.0034}_{-0.0027}$& $0.0017^{+0.0002}_{-0.0002}$& $0.26^{+0.04}_{-0.04}$\\
I Zw 18 NW & $20610^{+2450}_{-2240}$& $1.34^{+0.26}_{-0.53}$& $0.111^{+0.032}_{-0.036}$& $-0.19^{+0.43}_{-0.40}$& $0.16^{+0.05}_{-0.05}$& $-0.79^{+0.96}_{-0.98}$& $-6.5^{+2.4}_{-2.4}$& $0.0671^{+0.0032}_{-0.0036}$& $0.0032^{+0.0001}_{-0.0001}$& $0.40^{+0.23}_{-0.28}$\\
J1201+0211 & $17560^{+2670}_{-2540}$& $2.93^{+0.10}_{-0.10}$& $0.272^{+0.038}_{-0.059}$& $-5.63^{+1.39}_{-1.31}$& $0.08^{+0.52}_{-0.47}$& $2.20^{+0.76}_{-0.63}$& $-5.3^{+2.5}_{-3.2}$& $0.0668^{+0.0080}_{-0.0064}$& $0.0009^{+0.0003}_{-0.0003}$& $0.85^{+0.09}_{-0.10}$\\
J1119+5130 & $14970^{+2130}_{-1870}$& $-1.33^{+1.79}_{-1.85}$& $0.178^{+0.035}_{-0.040}$& $-0.35^{+0.36}_{-0.34}$& $0.17^{+0.13}_{-0.12}$& $-0.68^{+0.90}_{-0.89}$& $-5.9^{+3.0}_{-2.7}$& $0.0794^{+0.0043}_{-0.0040}$& $0.0017^{+0.0005}_{-0.0005}$& $0.55 ^{+0.08}_{-0.09}$\\
J1234+3901 & $21810^{+2840}_{-2870}$& $2.43^{+0.24}_{-0.37}$& $0.153^{+0.066}_{-0.090}$& $-3.58^{+3.48}_{-3.48}$& $-0.01^{+2.80}_{-2.55}$& $4.14^{+3.14}_{-2.80}$& $-6.1^{+2.6}_{-2.6}$& $0.0770^{+0.0200}_{-0.0167}$& $0.0034^{+0.0015}_{-0.0015}$& $-0.65^{+0.52}_{-0.62}$\\
J0133+1342$^a$ & $18030^{+2260}_{-2310}$& $2.28^{+0.16}_{-0.15}$& $0.382^{+0.035}_{-0.039}$& $-0.58^{+1.56}_{-1.58}$& $1.26^{+0.88}_{-0.81}$& $-1.81^{+0.77}_{-0.74}$& $-6.4^{+2.5}_{-2.5}$& $0.0765^{+0.0066}_{-0.0056}$& $0.0012^{+0.0003}_{-0.0003}$& $0.40^{+0.07}_{-0.07}$\\
J0825+3532$^a$ & $18760^{+2430}_{-2380}$& $2.24^{+0.14}_{-0.17}$& $0.374^{+0.037}_{-0.141}$& $-5.16^{+2.29}_{-1.62}$& $-1.66^{+0.55}_{-0.47}$& $0.03^{+0.74}_{-0.69}$& $-4.9^{+2.4}_{-3.4}$& $0.0544^{+0.0142}_{-0.0048}$& $\cdots$& $0.45^{+0.07}_{-0.09}$\\
J0125+0759$^a$ & $21610^{+1090}_{-1410}$& $1.89^{+0.13}_{-0.17}$& $0.353^{+0.016}_{-0.017}$& $18.53^{+2.80}_{-2.64}$& $8.00^{+1.26}_{-1.25}$& $-0.74^{+0.49}_{-0.47}$& $-6.8^{+2.1}_{-2.2}$& $0.0945^{+0.0045}_{-0.0043}$& $0.0010^{+0.0003}_{-0.0003}$& $0.16^{+0.10}_{-0.11}$\\
J0935-0115$^a$ & $19200^{+2710}_{-2280}$& $2.57^{+0.14}_{-0.15}$& $0.199^{+0.013}_{-0.015}$& $7.48^{+1.08}_{-1.04}$& $0.92^{+0.25}_{-0.26}$& $-0.81^{+0.46}_{-0.46}$& $-6.6^{+2.3}_{-2.3}$& $0.0667^{+0.0032}_{-0.0035}$& $0.0020^{+0.0002}_{-0.0002}$& $0.24^{+0.08}_{-0.09}$\\
\enddata
\tablecomments{(1): ID. (2)-(9): The properties determined with the YMCMC code (Section \ref{sec:y+}). (10): The doubly ionized He abundance (Section \ref{sec:y++}). (11): The radiation softness parameter (Section \ref{sec:yn}).\\$^a$ The galaxies that dose not meet the $\chi^2_\mathrm{YMCMC}$ criterion (Section \ref{sec:y+}).}
\end{deluxetable*}

\subsubsection{Neutral He Abundances}\label{sec:yn}
The contributions from $y^0$ can be estimated from the hardness of the ionizing radiation with the radiation softness parameter, $\eta_{\mathrm{rds}}$ \citep{vilchez1988}, defined as
\begin{equation}\label{eq:eta}
    \eta_\mathrm{rds} = \frac{\mathrm{O}^+}{\mathrm{S}^+}\frac{\mathrm{S}^{++}}{\mathrm{O}^{++}}.
\end{equation}
We calculate the $\mathrm{O}^+$, $\mathrm{O}^{++}$, $\mathrm{S}^+$, and $\mathrm{S}^{++}$ 
with the emission lines of [{\sc Oii}]$\lambda\lambda7320, 7330$, [{\sc Oiii}]$\lambda5007$, [{\sc Sii}]$\lambda\lambda6717, 6731$, and [{\sc Siii}]$\lambda9069$ for the Subaru galaxies \citep{dors2016}.
Table \ref{tab:ymcmc} lists the $\eta_\mathrm{rds}$ values of the Subaru galaxies. All of the Subaru galaxies have $\log \eta_\mathrm{rds} \lesssim 0.9$. Because \citet{pagel1992} find that the abundance of neutral helium is negligible for a galaxy with $\log \eta_\mathrm{rds} \lesssim 0.9$, we conclude $y^0$ is negligible for the Subaru galaxies.

\subsection{O Abundance}
We use the oxygen abundance reported in previous studies, which are measured by the direct method. Table \ref{tab:he-o} lists  
the oxygen abundances of the Subaru galaxies and their references.
\section{Results}\label{sec:yp}
\subsection{Primordial He Abundance}\label{sec:yp-regression}
\begin{figure*}
\gridline{\fig{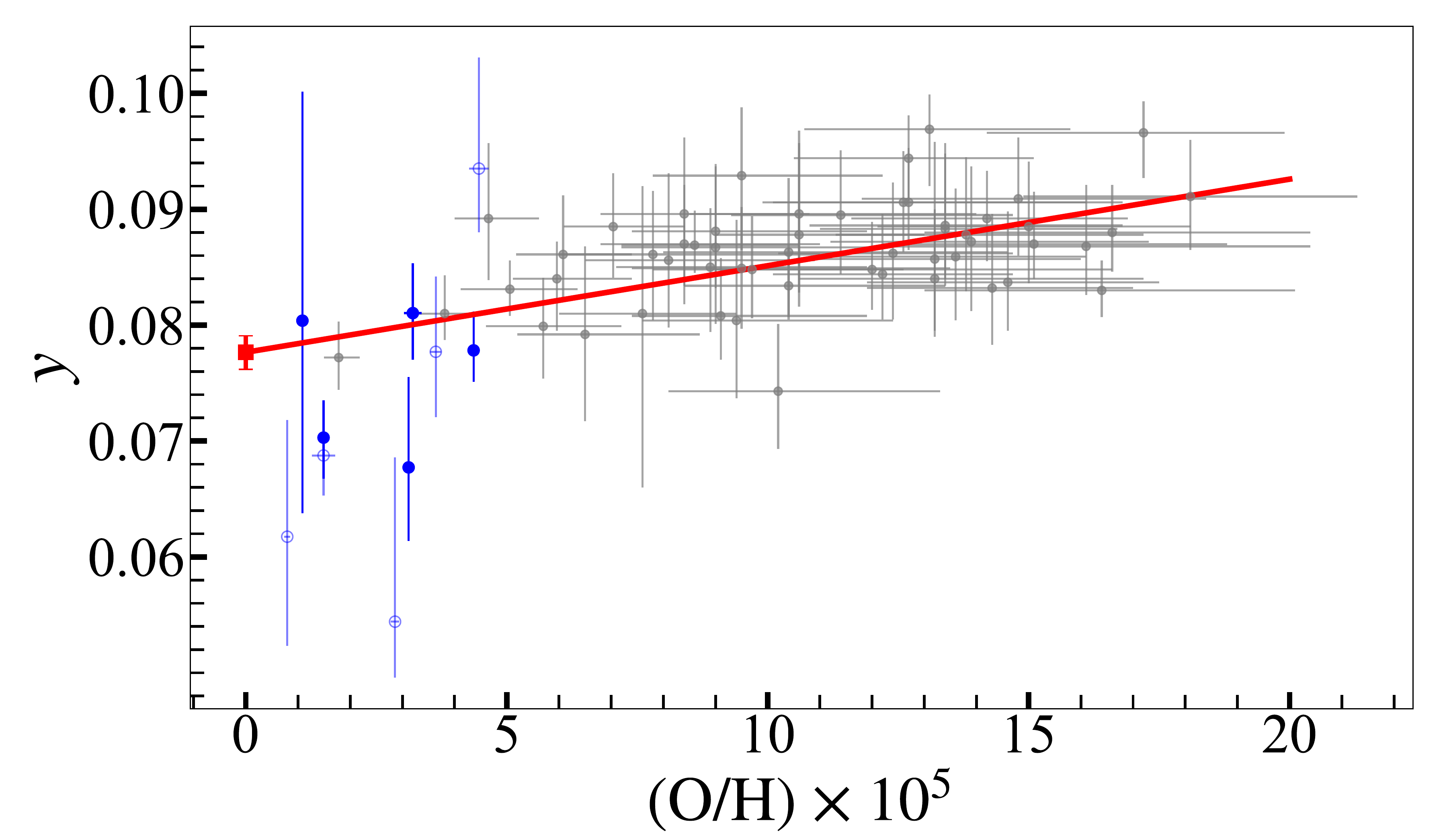}{1\textwidth}{}}
\caption{
Fifty nine galaxies (filled blue and gray circles) of our sample and the 5 EMPGs (open blue circles) that are excluded from the sample (Section \ref{sec:y+})
on the $y-\mathrm{O/H}$ plane. The blue (gray) circles represent the Subaru galaxies (the literature galaxies), which are described in Section \ref{sec:our_taget_object} (\ref{sec:objects_literature}). The red solid line shows the linear regression for the 59 galaxies and the red square denotes the $y_\mathrm{p}$ value determined by the linear regression.
\label{fig:regression}}
\end{figure*}

In the framework of the Big Bang cosmology, 
the helium element is produced by BBN and galactic chemical enrichment, while virtually no oxygen is created in the BBN.
\citet{peimbert1974,peimbert1976} have proposed
to determine $Y_\mathrm{P}$ with helium and oxygen abundance measurements by the linear regression of the form
%
\begin{equation}\label{eq:Y_linear}
    Y = Y_\mathrm{P}+\frac{\mathrm{d}Y}{\mathrm{d(O/H)}}\mathrm{(O/H)},
\end{equation}
where $Y$ is the helium mass fraction of a galaxy. %
$Y$ is derived with the equations
\begin{eqnarray}\label{eq:y_to_Y}
    Y &=& \frac{4y}{1+4y}(1-Z)\ \ \ \ \  {\rm and}\\
    Z &=& c\times \mathrm{(O/H)},
\end{eqnarray}
where $Z$ and $c$ are the heavy-element mass fraction and the coefficient, respectively.
%
%
%
Because $c$ is uncertain, the $y$ values cannot be converted to the $Y$ values precisely. To avoid the uncertainty, \citet{hsyu2020} have derived the primordial helium number abundance ratio $y_\mathrm{P}$ in the $y-\mathrm{(O/H)}$ plane by the linear regression of the form
%
\begin{equation}\label{eq:y_linear}
    y = y_\mathrm{P}+\frac{\mathrm{d}y}{\mathrm{d(O/H)}}\mathrm{(O/H)}.
\end{equation}
The likelihood function of their linear model does not contain terms corresponding to the uncertainties of O/H measurements, while \citet{hsyu2020} change the likelihood function with new values of O/H from Gaussian distributions with the mean values of observed values and the standard deviations of their errors at each step of MCMC sampling. 
In our study, to account for the uncertainties of O/H measurements in the same way as $y$ uncertainties, we consider the probability of obtaining the O/H measurements arising from `true' values. We maximize the log-likelihood function given by
\begin{align}\label{eq:log_likeli}
\begin{split}
    \log\left(\mathcal{L}\right) = -\frac{1}{2}\sum_{i}&\left[\frac{\left(y_i - a\mathrm{\left(\frac{O}{H}\right)}_i-b\right)^2}{\sigma_{y_i}^2 + a^2\sigma_{\mathrm{(O/H)}_i}^2+\sigma_\mathrm{int}^2} \right.\\
    & \left.+ \log\left(\sigma_{y_i}^2 + a^2\sigma_{\mathrm{(O/H)}_i}^2+\sigma_\mathrm{int}^2\right)\right],
\end{split}
\end{align}
with the slope $a\equiv \mathrm{d}y/\mathrm{d(O/H)}$ and
the primordial helium number abundance ratio $b\equiv y_\mathrm{P}$, and the intrinsic dispersion $\sigma_\mathrm{int}$ that is introduced 
for capturing unrecognized systematics of measurements \citep{cooke2018, hsyu2020}. Here, $y_i~(\sigma_{y_i})$ and $\mathrm{\left(\frac{O}{H}\right)}_i~(\sigma_{\mathrm{(O/H})_i})$ are the measured $y$ values (errors) and O/H values (errors), respectively. The summation of Equation (\ref{eq:log_likeli}) is over all galaxies in the sample. The result for our sample of the 64 galaxies is shown in Figure \ref{fig:regression}.
The regression yields

\begin{equation}\label{eq:yp_result}
\begin{split}
    y_\mathrm{P}=0.0777^{+0.0015}_{-0.0014},\\
    \frac{\mathrm{d}y}{\mathrm{d(O/H)}} = 75^{+15}_{-14},\\
    \sigma_\mathrm{int} \leq 0.0019~(95 \%).
\end{split}
\end{equation}
Note that we quote a $2\sigma$ upper limit on $\sigma_\mathrm{int}$ because it is consistent with zero.
Converting our $y_\mathrm{P}$ value to the mass fraction $Y_\mathrm{P}$ via $Y_\mathrm{P}=4y_\mathrm{P}/(1+4y_\mathrm{P})$, we obtain

\begin{equation}\label{eq:Yp}
    Y_\mathrm{P}=0.2370^{+0.0034}_{-0.0033}.
\end{equation}

\begin{figure*}
\gridline{\fig{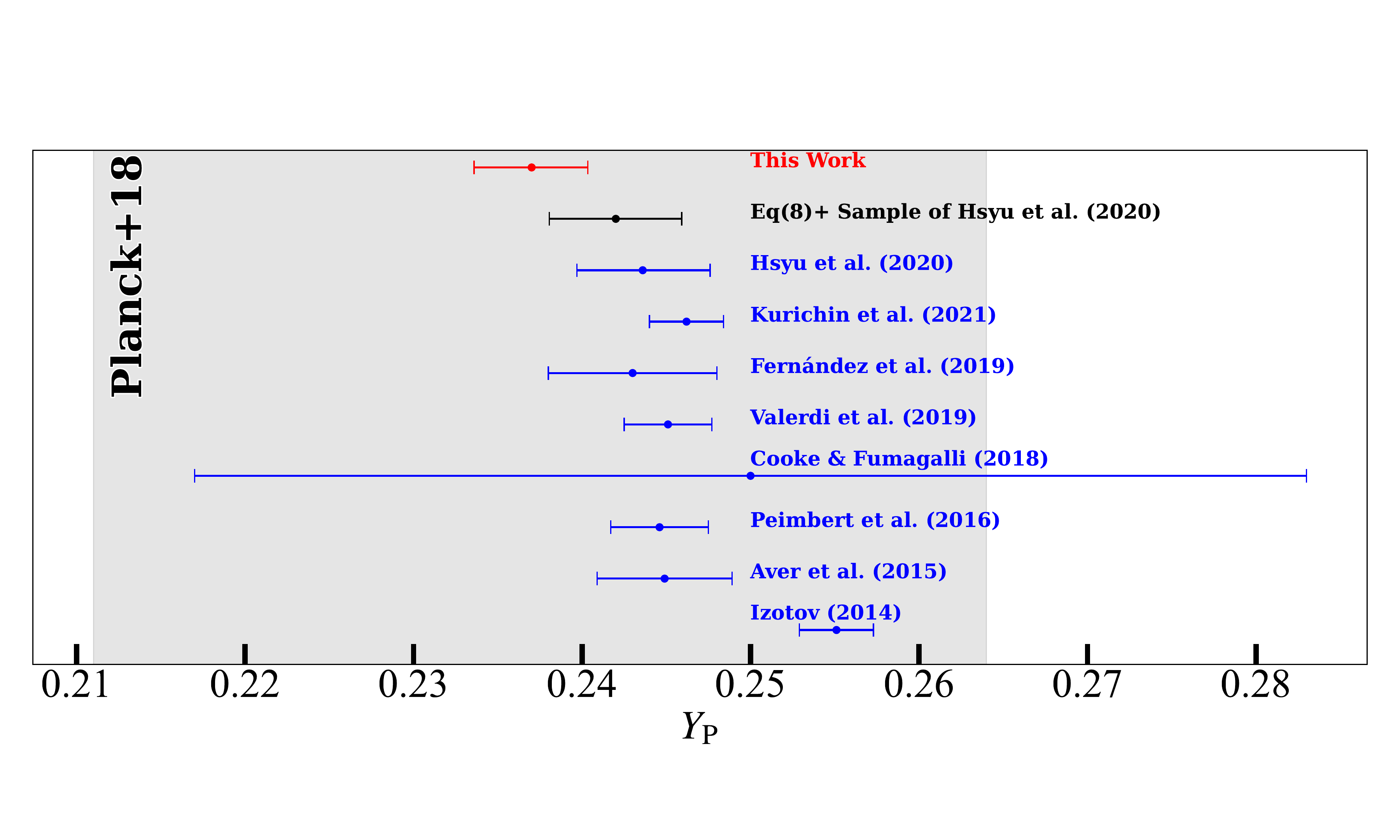}{1\textwidth}{}}
\caption{
Comparison of our $Y_\mathrm{P}$ values with those reported in recent literature. The blue circles with errors show the $1\sigma$ confidence regions derived from He emission line analyses \citep{izotov2014,aver2015,peimbert2016,fernandez2019,valerdi2019,hsyu2020,kurichin2021} and an analysis of a quasar absorption system \citep{cooke-fumagalli2018}. The gray region shows the constraint from CMB observations with $2\sigma$ confidence limit \citep{planck+18}. The red circle represents our result with the $1\sigma$ limit. The result with the sample from \citet{hsyu2020} using our likelihood function given by Equation (\ref{eq:log_likeli}) is denoted with the black circle.  
\label{fig:yp_comp}}
\end{figure*}

We compare the $Y_\mathrm{P}$ measurement of our study with those of previous studies in Figure \ref{fig:yp_comp}. 
Our $Y_\mathrm{P}$ measurement is comparable with those obtained by the method similar to ours
%
%
\citep{aver2015,peimbert2016,  fernandez2019,valerdi2019,hsyu2020,kurichin2021}.
However, our measurement is lower than the previous measurements at the $\sim 1 \sigma$ level. 

To explore the source of the $\sim 1 \sigma$-level difference, we apply our linear-regression method of Equation (\ref{eq:log_likeli}) to the sample of \citet{hsyu2020}, and present the obtained $Y_\mathrm{P}$ value in Figure \ref{fig:yp_comp} together with the one derived by \citet{hsyu2020}.
Although the linear-regression method of \citet{hsyu2020}
is different from our method, we confirm that our and \citeauthor{hsyu2020}'s results are almost identical, albeit with a negligibly small difference produced by the linear-regression methods. We also derive the $Y_\mathrm{P}$ value with the line fluxes obtain by the integration method to test whether the difference in the optical flux measuring methods affects our result. In this case,  with the 4 Subaru galaxies that meet the qualification criterion (see Section \ref{sec:y+}) and the literature galaxies, we obtained $Y_\mathrm{P} = 0.2373^{+0.0035}_{-0.0034}$. This value is almost the same as the one in Equation (\ref{eq:Yp}). We therefore conclude that our choice of the flux measurement method makes almost no difference in our result.

Because the main difference between our study and \citet{hsyu2020} is the inclusion of the EMPGs, we conclude that the source of the $\sim 1 \sigma$-level difference is the EMPGs covering the metal-poor end (i.e., small O/H) that is key for the $Y_\mathrm{P}$ determination (Figure \ref{fig:regression}).

Our $Y_\mathrm{P}$ value is in agreement with the one inferred from the CMB measurements \citep{planck+18} as well as the analysis of an absorption system in near-pristine intergalactic gas clouds 
along the light of a background quasar 
\citep{cooke-fumagalli2018}.


\begin{figure*}
\gridline{\fig{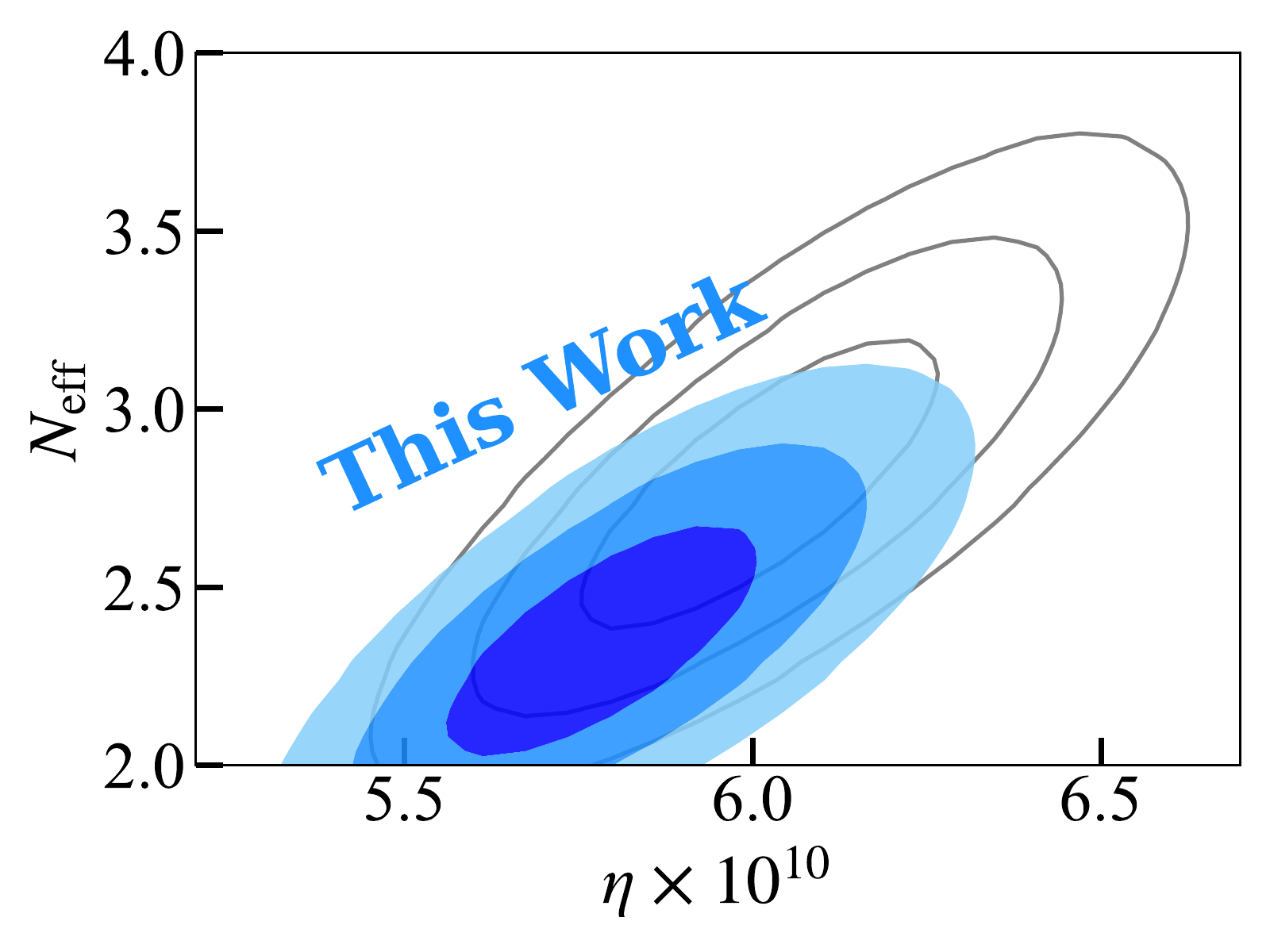}{1\textwidth}{}}
\caption{Comparison of our constraints on $N_\mathrm{eff}$ and $\eta$ (blue contours) with those of \citet{hsyu2020} (gray contours). These contours show $1\sigma, 2\sigma$, and $3\sigma$ confidence regions.
\label{fig:cont_comp}}
\end{figure*}

\subsection{Constraint on $N_\mathrm{eff}$}
The $Y_\mathrm{P}$ value provides powerful constrains on the cosmological parameters.
In the framework of the standard BBN model, $Y_\mathrm{P}$ strongly depends both on the baryon to photon ratio $\eta$ and the $N_\mathrm{eff}$ value.
%
%
%
We constrain $\eta$ and $N_\mathrm{eff}$ with our $Y_\mathrm{P}$ measurement and the primordial
deuterium abundance $D_\mathrm{P}$ measurement of
%
$D_\mathrm{P} =(2.527\pm0.030)\times10^{-5}$ \citep{cooke2018} 
by minimizing
\begin{multline}\label{eq:chi}
    \chi^2(\eta, N_\mathrm{eff}) = \frac{\left(Y_\mathrm{P, obs}-Y_{\mathrm{P, mod}}(\eta, N_\mathrm{eff})\right)^2}{\sigma^2_{Y_\mathrm{P,obs}}+\sigma^2_{Y_\mathrm{P,mod}}}\\
    +\frac{\left(\mathrm{D_{P, obs}}-\mathrm{D_{P, mod}}(\eta, N_\mathrm{eff})\right)^2}{\sigma^2_\mathrm{D_P,obs} + \sigma^2_\mathrm{D_P,mod}},
\end{multline}
with respect to $N_\mathrm{eff}$ and $\eta$, where the subscripts $\mathrm{obs}$ and $\mathrm{mod}$ denote the observational values and the theoretical BBN model values, respectively. To calculate the $Y_\mathrm{P,mod}$ and $D_\mathrm{P,mod}$ values for given values of $N_\mathrm{eff}$ and $\eta$, we use the version 3.0 of the PArthENoPE BBN code \citep{parthenope30}, fixing all input parameters of PArthENoPE except $N_\mathrm{eff}$ and $\eta$ to the standard values. In the calculation of $D_\mathrm{P,mod}$ and $Y_\mathrm{P,mod}$, we use the neutron lifetime $\tau_n = 879.4 \pm 0.6~\mathrm{s}$ \citep{pdg2020} and the relevant nuclear reaction rates from \citet{pisanti2021}. The errors of the $\tau_n$ and the nuclear reaction rates propagate to the errors of $D_\mathrm{P,mod}$ and $Y_\mathrm{P,mod}$. In Equation (\ref{eq:chi}), the $\sigma_\mathrm{D_P,mod}^2=(0.06)^2\times10^{-10}$ is the error of $D_\mathrm{P,mod}$ due to the uncertainty of the nuclear reaction rates and the $\sigma_{Y_\mathrm{P,mod}}^2=(0.00003)^2+(0.00012)^2$ is the error of $Y_\mathrm{P,mod}$, where the two terms correspond to the uncertainties of the nuclear reaction rates and the $\tau_n$, respectively \citep{parthenope30}.
We find 
\begin{eqnarray}\label{eq:2-neff}
    N_\mathrm{eff}=2.37^{+0.19}_{-0.24},\\
    \eta\times10^{10} = 5.80^{+0.13}_{-0.16}.
\end{eqnarray}

Figure \ref{fig:cont_comp} presents our constraint on $\eta$ and $N_\mathrm{eff}$, and comparison with the result of \citet{hsyu2020}. Our constraint is consistent with the one of \citet{hsyu2020} within the $1\sigma$ errors, while the best-estimate values of ours are slightly smaller than those of \citet{hsyu2020}. 


%

\section{Discussion}\label{sec:discussion}
\begin{figure*}
\gridline{\fig{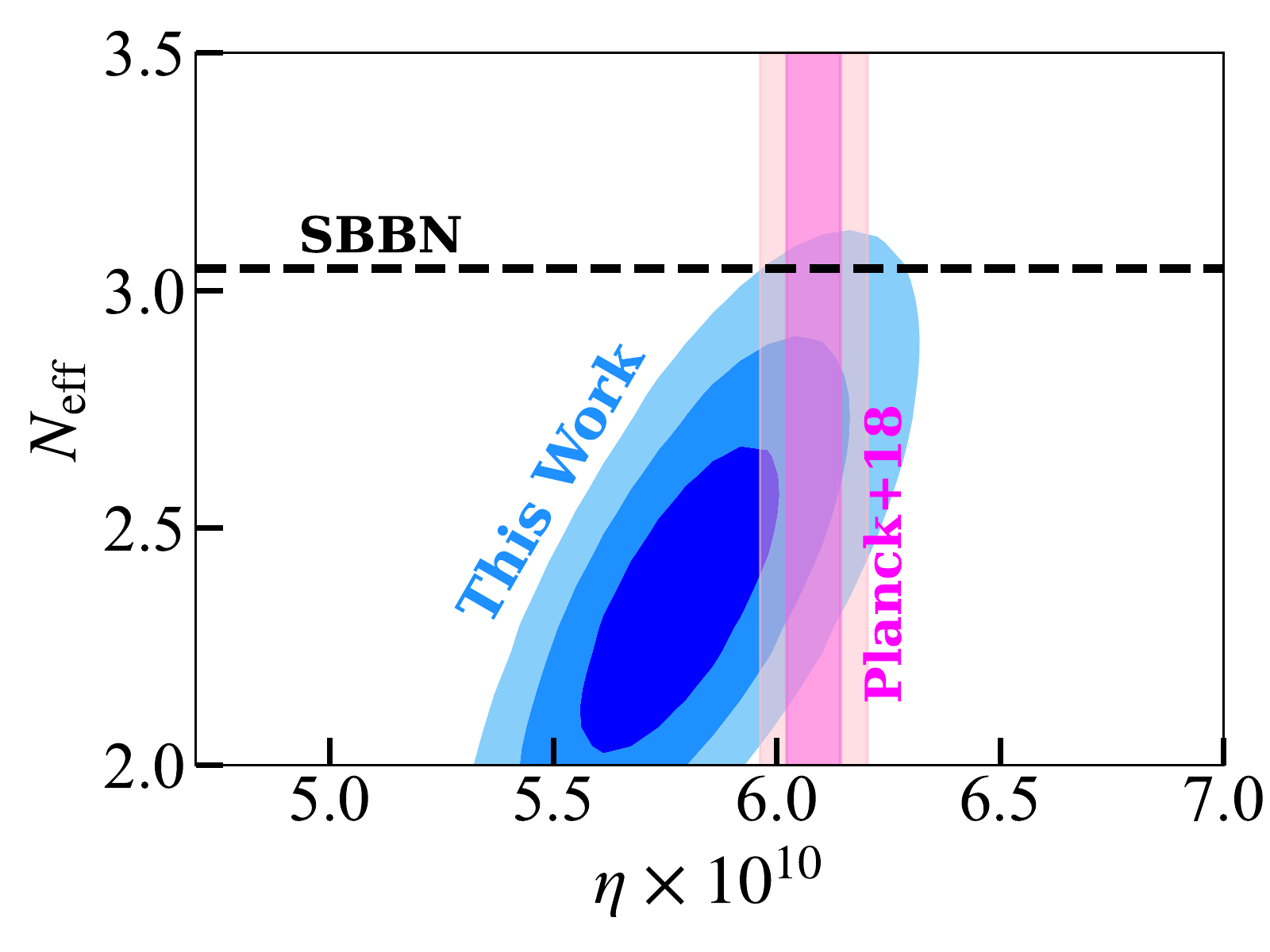}{1\textwidth}{}}
\caption{Observational constrains on $\eta$ and $N_\mathrm{eff}$. The blue contours show the $1\sigma$, $2\sigma$, and $3\sigma$ levels determined by this work. The black dashed line shows the standard model value of $N_\mathrm{eff}=3.046$. The magenta and light magenta bands represent the Planck constraint on $\eta$ at the $1\sigma$ and $2\sigma$ levels, respectively \citep{planck+18}. 
\label{fig:eta-neff}}
\end{figure*}

If the $N_\mathrm{eff}$ becomes smaller, the values of $Y_\mathrm{P}$ and $D_\mathrm{P}$ decrease. This is because the $\beta$ equilibrium between neutrons and protons continues for longer time reducing the abundance of neutrons, which are processed into light elements during the BBN. On the other hand, the smaller the $\eta$ gets, the larger $D_\mathrm{P}$ gets because the reactions which deplete deuterium become inefficient. Therefore, our smaller value of $Y_\mathrm{P}$ leads to the smaller values of $N_\mathrm{eff}$ and $\eta$. Figure \ref{fig:eta-neff} presents the constraint on $\eta$ and $N_\mathrm{eff}$, together with the one on $\eta$ obtained by \citet{planck+18}.
Our constraints suggest that there is a potential $\gtrsim 2\sigma$ tension with the Standard Model that predicts  
$N_\mathrm{eff}=3.046$ (Figure \ref{fig:eta-neff}). Moreover, our constraints agree with the Planck measurement in $\eta$ only at the $1-2\sigma$ level. 
%
%
This may be a hint of an electron-neutrino $\nu_e$ to anti-electron neutrino $\bar{\nu}_e$ asymmetry (i.e., lepton asymmetry), because the $\nu_e - \bar{\nu}_e$ asymmetry shifts the beta equilibrium between protons and neutrons before BBN, which changes the primordial element abundances. If $\nu_e$ increases (decrease), the primordial element abundances decrease (increase).
%
%
The $\nu_e - \bar{\nu}_e$ asymmetry
is represented by the degeneracy parameter of electron-neutrino, $\xi_\mathrm{e}\equiv\mu_{\nu_e}/T_{\nu_e}$ in natural units, where $\mu_{\nu_e}$ and $T_{\nu_\mathrm{e}}$ are the chemical potential and the temperature of $\nu_e$, respectively.
%
%
%
Here the $\xi_e$ can be both negative and positive and the $\nu_e - \bar{\nu}_e$ asymmetry is given by $n_{\nu_e} - n_{\bar{\nu}_e} \propto (\pi^2\xi_e+\xi^3_e)T_{\nu_e}^3$ with the Fermi-Dirac distribution function, where $n_{\nu_e}$ ($n_{\bar{\nu}_e}$) is the number density of (anti-)electron neutrinos.
%
%
Although the standard cosmology assumes $\xi_e=0$, so far whether this assumption is true is not revealed by the Standard Model nor astronomical observations \citep[e.g.,][]{kohri1997,popa2008,caramete2014,nunes2017,isabel2017}. Our low $Y_\mathrm{P}$ value (Figure \ref{fig:yp_comp}) may imply $\xi_e > 0$ \citep{sato1998, kohri1997}, while there are other possibilities \citep[e.g.,][]{khori2022}.
%
%

\begin{figure*}
\gridline{\fig{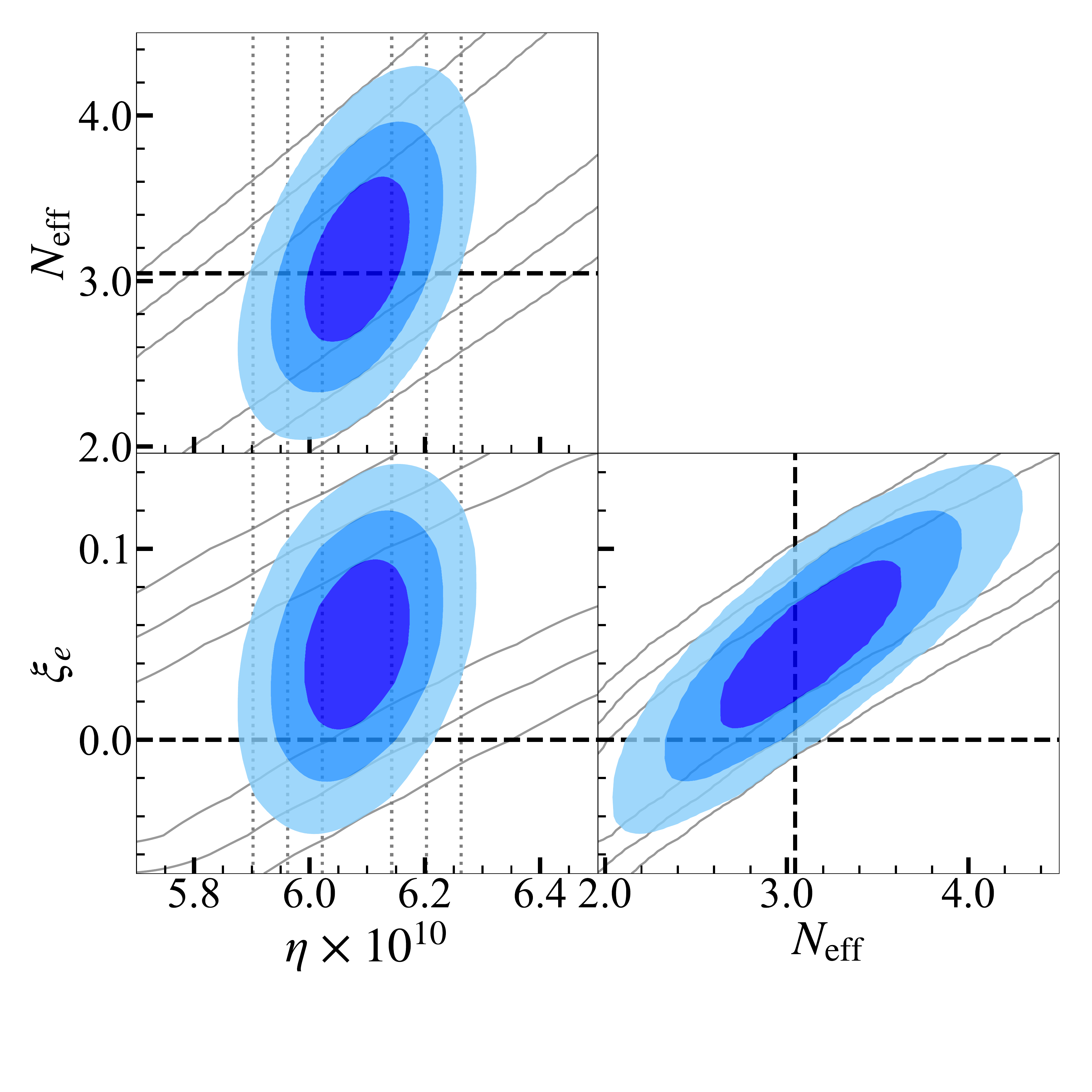}{1\textwidth}{}}
\caption{Constraints on $N_\mathrm{eff}, \eta$, and $\xi_\mathrm{e}$. The solid gray contours show the constraint with our $Y_\mathrm{P}$ value and the $D_\mathrm{P}$ measurement \citep{cooke2018}. The vertical dotted lines represent the \citet{planck+18} constraint on $\eta$. The constraint combining with the $Y_\mathrm{P}$, $D_\mathrm{P}$, and $\eta$ measurements is shown with blue contours. These contours show $1\sigma, 2\sigma$ and $3\sigma$ confidence limits.
The standard model values of $N_\mathrm{eff}=3.046$ and $\xi_e=0$ are represented with black dashed lines.
\label{fig:3-men}}
\end{figure*}

To constrain $\xi_e$ as well as $N_\mathrm{eff}$ and $\eta$, we minimize 
\begin{multline}\label{eq:chi_3}
    \chi^2(\eta, N_\mathrm{eff}, \xi_e) = \frac{\left(Y_\mathrm{P, obs}-Y_{\mathrm{P, mod}}(\eta, N_\mathrm{eff}, \xi_e)\right)^2}{\sigma^2_{Y_\mathrm{P,obs}}+\sigma^2_{Y_\mathrm{P,mod}}}\\
    +\frac{\left(\mathrm{D_{P, obs}}-\mathrm{D_{P, mod}}(\eta, N_\mathrm{eff}, \xi_e)\right)^2}{\sigma^2_\mathrm{D_P,obs} + \sigma^2_\mathrm{D_P,mod}}
    +\frac{(\eta-6.132)^2}{0.038^2},
\end{multline}
allowing $\xi_e$, $N_\mathrm{eff}$, and $\eta$ to vary independently of each other as input parameters of PArthENoPE.
In the equation (\ref{eq:chi_3}), in order to break the degeneracy between the parameters, we impose a Gaussian prior of $\eta\times10^{10}=6.132\pm0.038$, which comes from the marginalized constraint on the baryon density by \citet{planck+18} where $N_\mathrm{eff}$ and $Y_\mathrm{P}$ are treated as free parameters.
Figure \ref{fig:3-men} presents 2-dimensional marginalized constraints on the three parameters of $\xi_e$, $N_\mathrm{eff}$, and $\eta$. The gray contours show the constraint obtained without the prior of eta, illustrating a degeneracy between the three parameters. The vertical dotted lines correspond to the Planck measurement of $\eta$. In the left two panels of Figure \ref{fig:3-men}, the gray and dotted contours intersect in a region of the parameter spaces. With the full combined results from the $Y_\mathrm{P}$, $D_\mathrm{P}$, and $\eta$ measurements, we break the parameter degeneracy, and find

\begin{eqnarray}\label{eq:3-men}
 N_\mathrm{eff}=3.11^{+0.34}_{-0.31},\\
 \eta\times10^{10}=6.08^{+0.06}_{-0.06},\\
 \xi_e=0.05^{+0.03}_{-0.02}.
\end{eqnarray}
The derived $\xi_e$ value is higher than 0 at the $\sim  2\sigma$ level. This may be a hint of 
the lepton asymmetry
with
an excess in the number of $\nu_e$ to that of $\bar{\nu}_e$. To realize the universe with $\xi_e \sim 0.05$, new physics for lepton number generation may be required \citep{kawasaki2022}.
%
%

As shown in the right panel of Figure \ref{fig:3-men}, there is a correlation between $\xi_e$ and $N_\mathrm{eff}$. This is because the effects of $N_\mathrm{eff}$ and $\xi_e$ on the BBN compensate for each other. A positive value of $\xi_e$ decreases the number of neutrons, which are in equilibrium with protons, while a $N_\mathrm{eff}$ value larger than $3.046$ ends the equilibrium at an earlier time, which means more neutrons are left before the BBN.
Our positive value of $\xi_e$ allows
$N_\mathrm{eff}$
significantly higher than the results in Equation (\ref{eq:2-neff}). 
%
%
While the $N_\mathrm{eff}$ value of our best estimate (Equation (\ref{eq:3-men})) is comparable with the one of the Standard Model ($N_\mathrm{eff}=3.046$), our best estimate could be as high as $N_\mathrm{eff}=3.45$, which can ameliorate the Hubble tension (Section \ref{sec:intro}), at the 68\% confidence level. 
Using the results of previous studies by the method similar to ours \citep{aver2015,peimbert2016,fernandez2019,valerdi2019,hsyu2020,kurichin2021}, we also confirm this trend of the central values of $\xi_e > 0$ and $N_\mathrm{eff}>3.046$ allowing for $N_\mathrm{eff}\simeq3.4$ at the 68\% confidence level. This trend is consistent with a cosmological model proposed by \citet{seto2021} to reduce the Hubble tension without spoiling BBN.
Because the contribution of $\xi_e \sim 0.05$ to increasing the $N_\mathrm{eff}$ value from 3.046 is small ($\lesssim \mathcal{O}(0.01)$) with the Fermi-Dirac distribution function, the existence of extra-radiation is necessary to realize $N_\mathrm{eff}\sim3.4$.
Although the errors of our measurements are still too large to conclude, there is a possibility of the $\nu_e - \bar{\nu}_e$ asymmetry and extra-radiation which may provide the high $N_\mathrm{eff}$ value that resolves the Hubble tension (Section \ref{sec:intro}). 
%
%

%
%


%

%



%


\section{Summary}\label{sec:sum}
Using Subaru/MOIRCS, IRCS, and SWIMS, we conducted NIR spectroscopic observations covering the He{\sc i}$\lambda$10830 line for 
the galaxies, which are classified as EMPGs with a metallicity less than 0.1 solar metallicity. 
Removing one EMPG with a clear signature of outflows,
we determine He abundances of the 10 $(=11-1)$ galaxies using the NIR spectroscopic data and the pre-existing optical spectroscopic data.
We explore the best-fit physical parameters of the nebulae with the observed line fluxes by the MCMC technique. 
Selecting the 5 EMPGs from the Subaru galaxies, whose physical properties are reliably determined, we construct a sample of a total of 59 galaxies consisting of our 5 EMPGs and 54 galaxies (including 3 EMPGs) taken from the literature, increasing the number of EMPGs from 3 to 8 $(=3+5)$ that are key for the primordial He abundance $Y_\mathrm{P}$ determination. 
We
%
%
%
derive $Y_\mathrm{P}$ with the 59 galaxies, and constrain the effective number of neutrino species $N_\mathrm{eff}$, the baryon to photon ratio $\eta$, and the electron-neutrino degeneracy parameter $\xi_e$.
Our main results are summarized below.
\begin{itemize}
    \item The linear regression $y-\mathrm{(O/H)}$ for the 59 galaxies gives $Y_\mathrm{P}=0.2370^{+0.0034}_{-0.0033}$. Our $Y_\mathrm{P}$ value is in agreement with the one inferred from the CMB measurements \citep{planck+18}, and comparable with those of the previous galaxy observations. 
    %
    \item With our $Y_\mathrm{P}$ value and the $D_\mathrm{P}$ measurement given by \citet{cooke2018}, we obtain $N_\mathrm{eff}=2.37^{+0.19}_{-0.24}$ and $\eta\times10^{10}=5.80^{+0.13}_{-0.16}$ by the $\chi^2$ minimization. The constraint on $N_\mathrm{eff}$ is in a potential $\gtrsim 2\sigma$ tension with the Standard Model predicting $N_\mathrm{eff}=3.046$. 
    
    \item Motivated by the potential tension, we allow a $\nu_e - \bar{\nu}_e$ asymmetry (i.e. non-zero $\xi_e$) for $N_\mathrm{eff}$ and $\eta$ constraints.
    %
    We obtain the best-fit parameters, $\xi_e = 0.05^{+0.03}_{-0.02}$, $N_\mathrm{eff}=3.11^{+0.34}_{-0.31}$, and $\eta\times10^{10}=6.08^{+0.06}_{-0.06}$, where the $N_\mathrm{eff}$ and $\eta$ values agree with the Standard Model and the Planck measurement, respectively.
    %
    Our constraints suggest a $\nu_e - \bar{\nu}_e$ asymmetry and allow for a high value of $N_\mathrm{eff}$ up to $N_\mathrm{eff}=3.45$ within the $1\sigma$ level, which may mitigate the Hubble tension.
\end{itemize}

\begin{acknowledgments}
We thank the anonymous referee for a careful reading and valuable comments that improve the clarity of the paper. We are grateful to Erik Aver, Kazunori Kohri, Oleg Kurichin, and J. Xavier Prochaska for giving us helpful comments and discussions. This work is supported by the World Premier International Research Center Initiative (WPI Initiative), MEXT, Japan, as well as KAKENHI Grant-in-Aid for Scientific Research (A)(20H00180, 21H04467, 21J00153, 20K14532, 21H04499, 21K03614, 21K03622, and 22H01259) through the Japan Society for the Promotion of Science (JSPS). This work
was supported by the joint research program of the Institute for Cosmic Ray Research (ICRR), University of
Tokyo.
\end{acknowledgments}

%

\vspace{5mm}

\software{corner.py \citep{corner}, emcee \citep{emcee}, Matplotlib \citep{matplotlib}, SciPy \citep{scipy}}

\bibliography{yp_paper}{}
\bibliographystyle{aasjournal}

\end{document}